\documentclass[apj]{emulateapj}
\shorttitle{Tidal  breakup  of binary stars at the Galactic Center-II}
\shortauthors{Antonini, Lombardi \& Merritt}
\newcommand\rsun{\rm\,R_\odot}
\newcommand\msun{\rm\,M_\odot}
\newcommand\kms{{\rm\,km\,s^{-1}}} 
\newcommand\au{{\rm\,au}} 
 
\begin{document}

\title{Tidal breakup of binary stars at the Galactic Center. II. \\
Hydrodynamic simulations}

\author{Fabio~Antonini}
\email{antonini@astro.rit.edu}
\affil{Department of Physics and Center for Computational Relativity and Gravitation, 
Rochester Institute of Technology, 85 Lomb Memorial
Drive, Rochester, NY 14623, USA}

\author{James~C.~Lombardi Jr.}
\affil{Department of Physics, Allegheny College, 520 North Main Street, Meadville, PA 16335, USA}

\author{David Merritt}
\affil{Department of Physics and Center for Computational Relativity and Gravitation, 
Rochester Institute of Technology, 85 Lomb Memorial
Drive, Rochester, NY 14623, USA}

\subjectheadings{black hole physics-Galaxy:center-Galaxy:kinematics and dynamic}

\begin{abstract}
In Paper I, we followed the evolution of binary stars as they
orbited near the supermassive black hole (SMBH) at the Galactic
center, noting the cases in which the two stars would come close
enough together to collide. In this paper we replace the point-mass 
stars by fluid realizations, and use a smoothed-particle hydrodynamics
(SPH) code to follow the close interactions.  We model the binary 
components as main-sequence stars with initial masses of 
1, 3 and 6 Solar masses, and with chemical composition profiles taken 
from stellar evolution codes.  Outcomes of the close interactions include
mergers, collisions that leave both stars intact, and ejection
of one star at high velocity accompanied by capture of the other
star into a tight orbit around the SMBH.  For the first time, we
follow the evolution of the collision products for many ($\gtrsim 100$) orbits
around the SMBH.  Stars that are initially too small to be tidally
disrupted by the SMBH can be puffed up by
close encounters or collisions, with the result that tidal stripping 
occurs in subsequent periapse passages. In these cases, mass loss 
occurs episodically, sometimes for hundreds of orbits before the star 
is completely disrupted.  Repeated tidal flares, of either increasing
or decreasing intensity, are a predicted consequence. In collisions 
involving a low-mass and a high-mass star, the merger product acquires
a high core hydrogen abundance from the smaller star, effectively 
resetting the nuclear evolution ``clock'' to a younger age.  Elements
like Li, Be and B that can exist only in the outermost envelope of
a star are severely depleted due to envelope ejection during collisions
and due to tidal forces from the SMBH. 
Tidal spin-up can occur due to either  a collision or  tidal torque by the SMBH 
at periapsis.
 However, in the absence of collisions,
tidal spin-up of stars is only important in a narrow range of periapse
distances, $r_{\rm t}/2\lesssim r_{\rm per} \lesssim r_{\rm t}$ with $r_{\rm t}$ the tidal
disruption radius. 
We discuss the implications of these results for
the formation of the S-stars and the hypervelocity stars.
\footnote{Movies can be found at  http://astrophysics.rit.edu/fantonini/tbbs2/} \end{abstract}

\section{Introduction}
Tidal breakup of binary stars by the supermassive black
hole (SMBH) at the Galactic center (GC) has been invoked 
to explain a number of otherwise puzzling  discoveries, 
including the hypervelocity stars (HVSs) that are observed
in the halo of the Milky Way \citep{b05,b06b,b07,b09}, and the S-stars, apparently
young, main-sequence stars in tight eccentric orbits around
the SMBH \citep{EI:05,gill09}.
As first pointed out by J. Hills,
close passage of a binary star near a SMBH can result in 
an exchange interaction, such that one component of the binary
is ejected with greater than escape velocity while the other
star is scattered onto a tightly-bound orbit \citep{HI:88,YU:03}.
The predictions of this model are broadly consistent with 
the observed properties of both the HVSs \citep{brom06} and the S-stars \citep{PA:09,PA:10}.

The origin of the binary progenitors of the HVSs is 
not clear. 
One possibility is that the binaries originated at distances
of a few parsecs from the GC and were subsequently scattered 
inward by  ``massive-perturbers''  \citep{PHA}.
In this scenario, most of the binaries will lie on either 
unbound or weakly-bound orbits with respect to the SMBH, 
and they will encounter it only once before being scattered 
onto different orbits.
Alternatively,  the binaries may form  closer to the SMBH, 
perhaps in the young (or an older)  stellar disk that is 
observed between $\sim 0.04$ pc and $\sim0.5$ pc from
the SMBH \citep{NC:05,P:06,NC:07,L:07}.
Also, \citet{P:09} suggested that binaries could be left near the GC
through a triple disruption by the SMBH.
In these latter cases, the binaries would  be bound to the SMBH and would
encounter it many times before being disrupted.

If the finite sizes of stars are taken into account,
a number of outcomes are possible in addition to simple 
binary disruption.
The two stars can collide, resulting in a merger
if the relative velocity is less than stellar escape
velocities  \citep{GL:07}.
Since the radius of tidal disruption of single stars
by the SMBH is comparable to the binary disruption radius,
stars can also be tidally disrupted by the SMBH, either before or
after their close interaction with each other.

In Paper I \citep{me:09} we presented the results of
a large number of $N$-body integrations of point-mass binary stars 
on eccentric orbits around the GC SMBH.
In many cases, the trajectories of the two stars were found to 
imply a physical collision, assuming that the unperturbed stars
had radii similar to those of normal main-sequence stars of the same mass.
The probability of physical encounters was found to increase
significantly if the binaries were allowed to complete many
orbits about the SMBH.
In some cases, one or both stars also passed close enough to the SMBH
that gravitational tides would be expected to significantly affect
their internal structure.

In this paper, we use smoothed-particle hydrodynamics  
(SPH) simulations to study the binaries from Paper I
that approached closely enough to physically interact.
The $N$-body simulations of Paper I were first used to identify
initial conditions that resulted in close interactions between
the two stars.
The point-mass stars were then realized as macroscopic, 
fluid-dynamical models and integrated forward in the gravitational
field of the SMBH using an SPH algorithm.
As in Paper I, we followed the trajectories 
for multiple orbits around the SMBH, allowing us, for
the first time, to investigate the consequences of repeated
tidal interactions with the SMBH.

In \S2 we briefly discuss time scales for binary disruption
at the GC.
Our  initial conditions and numerical methods are described
in \S 3 and the results in \S 4.
Some observable consequences are presented in \S 5.
\S 6 sums up.

\section{The survival time of binaries at the Galactic Center}

In a dense environment, binaries may evaporate due to dynamical  interactions 
with field stars if
\begin{equation}
|E|/({M_{\rm b} \sigma^2}) \lesssim 1,
\end {equation}
with $E$ the internal orbital energy of the binary,
$M_{\rm b}$ the binary mass,  and $\sigma$ the 
one-dimensional velocity dispersion  of the stellar background.
In principle, because most of the binaries at the GC are expected to be
``soft'', $|E| \lesssim M_{\rm b} \sigma^2$,
and because the binary evaporation time  $t_{\rm ev}$ is a  function 
of the distance from the SMBH, 
the variation of $t_{\rm ev}$ with galactocentric radius can 
be used to constrain the origin of the HVSs \citep{PER:09}. 
If the evaporation time at some radius is  shorter than the  
lifetime of a typical  main-sequence star, 
the  stellar population in this region would be 
dominated by isolated (i.e. single) stars.

Here we show that the survival time of binaries at galactocentric distances 
$r< 0.1$pc  is likely to be comparable to the typical 
main-sequence lifetimes of most stars in this region. 
We also show that, within a radius of $r \sim 0.3$pc, 
$t_{\rm ev}$ becomes essentially independent of radius.

\begin{figure}{~~}
\begin{center}
\includegraphics[angle=270,width=3in]{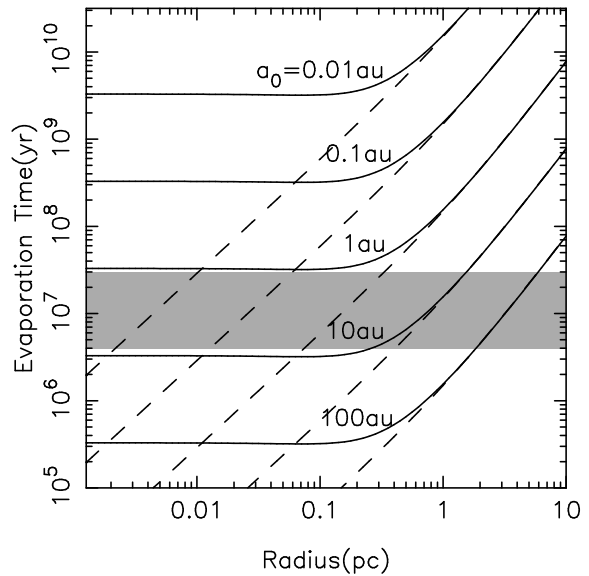} 
\caption{Evaporation time of binaries vs. galactocentric radius 
for different values of the binary semimajor-axis $a_0$. 
Solid curves show the evaporation time for the density model of 
equation~(\ref{den}) with $\gamma=0.5$ while the dashed curves 
correspond to the coreless model with slope  $\gamma=1.8$. 
The filled grey region gives the ages of the S-stars \citep{EI:05}.}
\label{ev}
\end{center}
\end{figure}

Beyond $\sim 1$ pc from Sgr A$^*$, the mass density determined from
the stellar kinematics follows 
$\rho \sim r^{-\beta}$, $1.5\lesssim\beta\lesssim 2$ \citep[e.g.][]{Oh:09}.
At smaller radii, {\it number counts} of the dominant (old) stellar 
population near the GC suggest a space density that is weakly rising,
or falling, toward the SMBH, inside a core of radius  $\sim 0.5$pc
\citep{BSE,D:09,B:10}.
Approximating the mass density as a broken power-law,
\begin{equation}
\label{den}
\rho(r)=\rho_0 \left(  \frac{r}{r_0} \right)^{-\gamma}
 \left[
1+\left( \frac{r}{r_0} \right)^2\right]^{(\gamma-\beta)/{2}}
\end{equation}
with $r_0=0.3$pc and $\beta =1.8$, and setting
$\rho_0=1.3 \times 10^6{\rm M_{\odot}pc^{-3}}$ gives a good
fit to the space density outside the core \citep[e.g.][]{M:10}.
At smaller radii, the uncertainties in $\rho$ are represented
by the poorly determined value of $\gamma$.

The evaporation time is given by \citep{BT} :
\begin{equation}
t_{\rm ev}=\frac{M_{\rm b}\sigma}{M~16\sqrt{\pi} \rho a_0 \ln\Lambda},
\end{equation}
where ${\rm ln} \Lambda$ is the Coulomb logarithm, $M$ the mass of the field stars, $a_0$
the binary semimajor-axis,  and $\sigma$ is  calculated from the Jeans equation,
\begin{equation}
\rho(r)\sigma(r)^2 = G\int_r^{\infty} dr' r'^{-2} \left[M_{\bullet}+M_\star(<r')\right]\rho(r'),
\label{jeans}
\end{equation} 
with $M_\star(<r)$ the total mass in stars  within $r$, and
$M_{\bullet}$  the mass of the central black hole.
Hereafter, we adopt $M_{\bullet}= 4\times 10^6\msun$ \citep{ghez08,gill09}.
In Figure~\ref{ev} we plot the evaporation time of binaries in the
density model of equation (\ref{den}) as a function of galactocentric radius, 
 assuming $M_{\rm b}=2M$, 
$\ln \Lambda=15$ and  two different values of the internal slope: 
$\gamma=0.5$ representative of the observed distribution   and 
 $\gamma=1.8$ which corresponds approximately   to a  relaxed system 
around a SMBH \citep{BW:76}.

\begin{figure}
  \begin{center}
    \includegraphics[width=84mm]{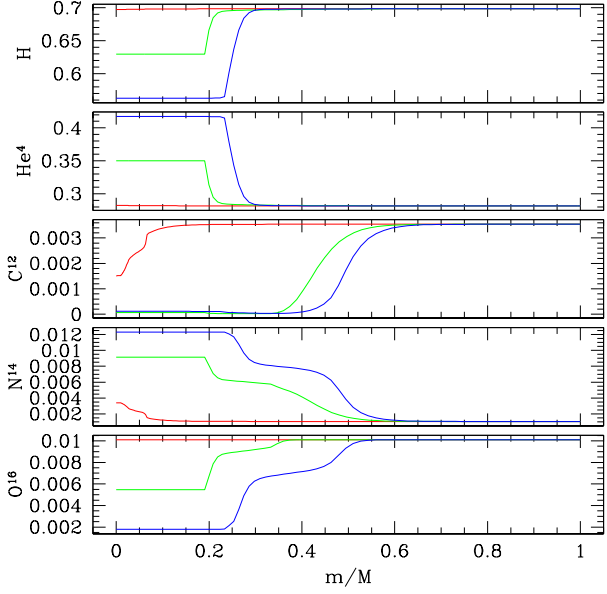}
  \end{center}
  \caption{Fractional chemical abundances (by mass) versus enclosed
  mass fraction $m/M$ for our $M=1 \msun$ (red curves), $3 \msun$ (green curves), and
  $6 \msun$ (blue curves) stars, as calculated by the TWIN stellar
  evolution code.}
  \label{fig:threeparents}
\end{figure}

\begin{figure}
  \begin{center}
    \includegraphics[width=84mm]{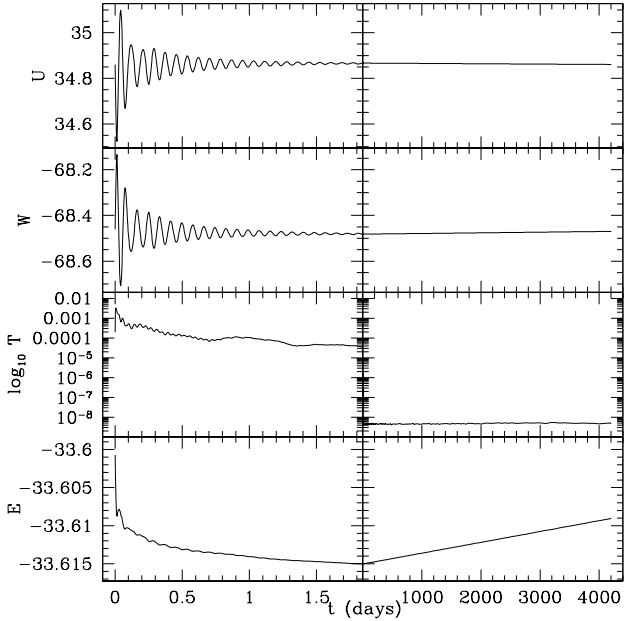}
  \end{center}
  \caption{
Internal energy $U$, gravitational potential energy $W$, kinetic
energy $T$, and total energy $E$ versus time $t$ for the relaxation
(left panels) and subsequent dynamical evolution in isolation (right
panels) of the SPH model for a $6 \msun$ star.  Note that the time $t$ is
shown on different linear scales for the relaxation and the
dynamical evolution.  Energies are in
units of $10^{48}$ erg.
}
  \label{fig:vlnb}
\end{figure}

\begin{figure}
  \begin{center}
    \includegraphics[width=84mm]{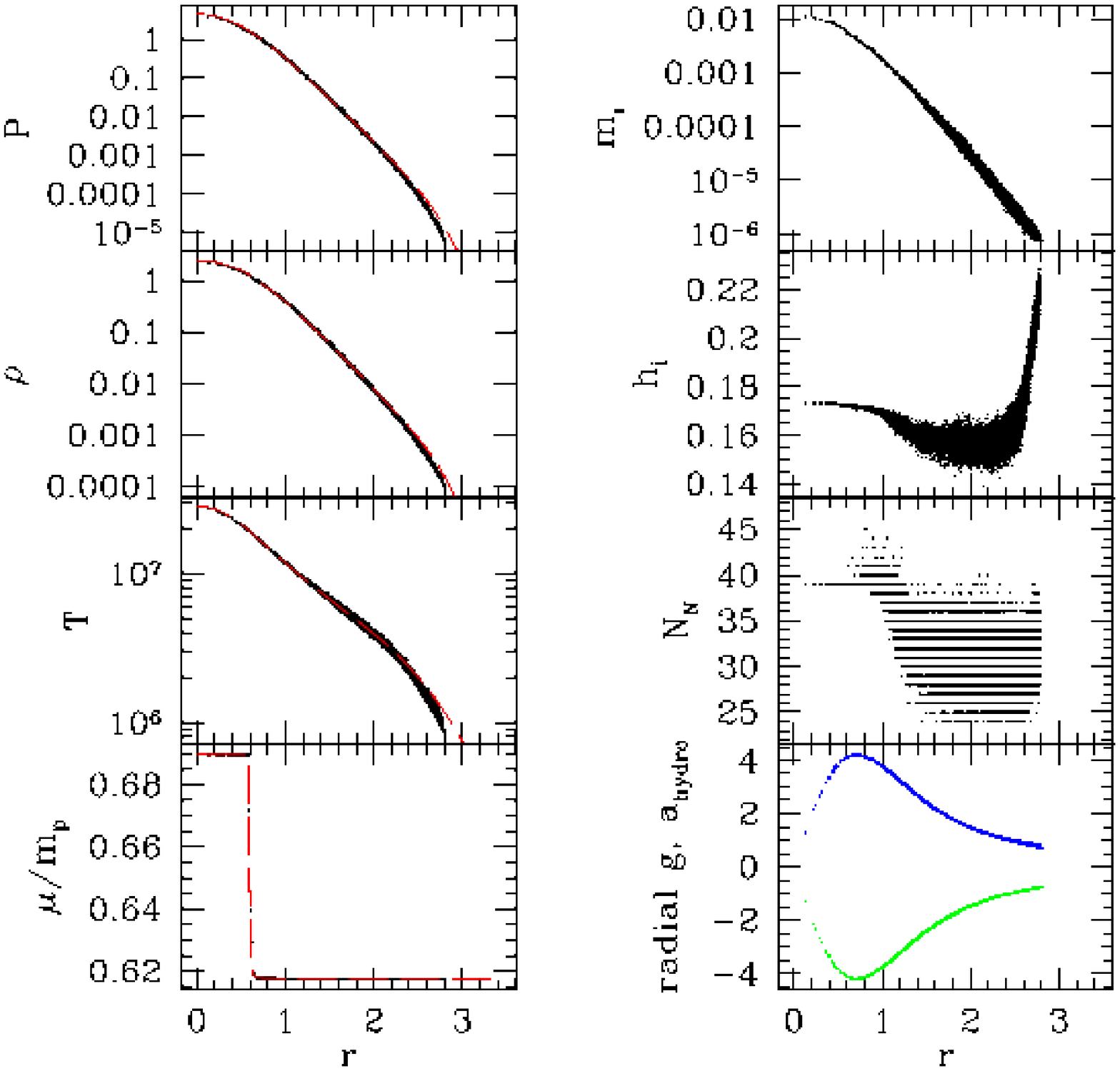}
    \includegraphics[width=84mm]{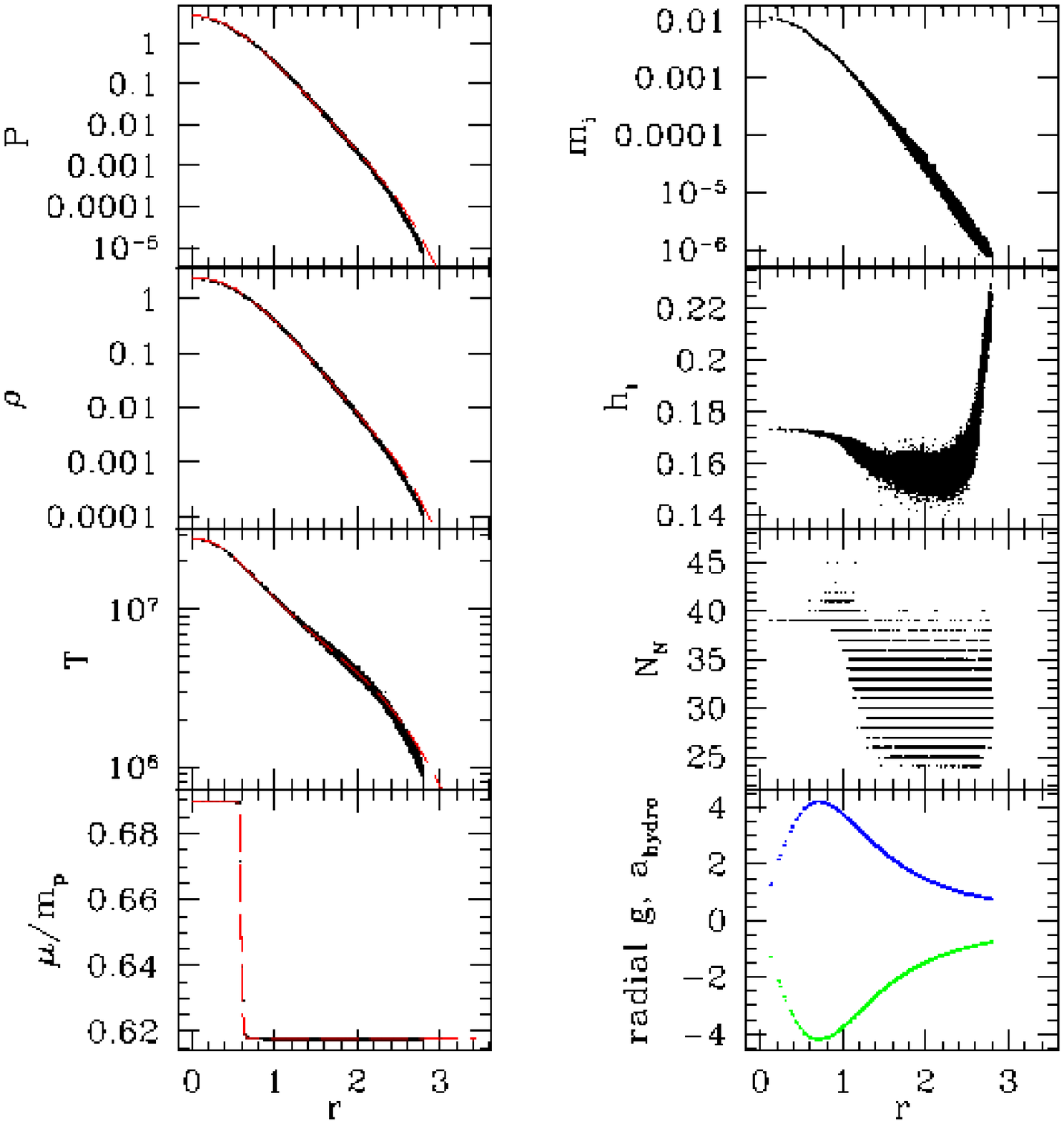}
  \end{center}
  \caption{Radial profiles of the SPH model for a $6 \msun$ star
    both at the end of relaxation (upper panels) and after 4200 days
    of hydrodynamical evolution (lower panels).    
    The frames in the left column show profiles of
    pressure $P$, density $\rho$, temperature $T$ (in Kelvin), and
    mean molecular weight $\mu$ in units of the proton mass $m_p$,
    with the dashed curve representing results the TWIN evolution code
    and dots representing particle data from our SPH model.  The right
    column provides additional SPH particle data: individual SPH
    particle mass $m_i$, smoothing length $h_i$, number of neighbors
    $N_N$, and radial component of the hydrodynamic acceleration
    $a_{\rm hydro}$ (upper data) and gravitational acceleration $g$
    (lower data).  Unless otherwise stated, quantities are in solar units ($G=\msun=\rsun=1$).
}
  \label{fig:plnc100_m6}
\end{figure}

From the figure it is clear that the survival time of binaries  
would be greatly increased if 
the distribution of stars  at the GC  has a low density core 
(for comparison, see also Figure 1 in Perets [2009]). 
The figure also shows that, at any radius, for $a_0 \lesssim 1\au$,
this time is larger than the typical lifetime of the  the S-stars.
Therefore, it cannot be excluded  that the S-stars were initially 
part of binary systems  originating at galactocentric distances  of 
few tens of milliparsecs. 

We finally  note  that, in the context of this paper,  
it might be more appropriate to compare 
 $t_{\rm ev}$ with the time scale required 
to drive the eccentricities  of the stars away from their initial
values to produce quasi-radial orbits. 
At the GC this time can be of the order of few Myr 
 \citep{LBK,MAD:08,MA:09,PA:09,Fujii:10}, which is  much shorter 
 than the  typical evaporation time scale of binaries with   
$a_0 \leq 0.2\au$. 

\section{Initial conditions and numerical method}

\subsection{Orbital initial conditions}

In Paper I we used the high-accuracy numerical integrator 
{\tt ARCHAIN} \citep{MA:02,MM:06}
to study the dynamics of main-sequence binary stars on highly elliptical  
orbits whose periapsides lay close to the SMBH.
We determined the final orbital properties of both ejected and bound stars.
Initial conditions consisted of  equal-mass binaries on circular relative
orbits with random  orientations, initial separations $a_0$ in the range
$0.05$ - $0.2\au$, 
and individual stellar masses $M$ of $3\msun-6\msun$.
The binaries were given a tangential initial velocity 
in the range $4$ km s$^{-1}$ - $85$ km s$^{-1}$ and initial 
distances of $d=0.01-0.1$ pc from the SMBH. 
%of mass$M_{\bullet}=4\times 10^6\msun$. 
%and then, the stars moved  initially on bound orbits
%with high eccentricity ($e \sim 0.95$).
Using the mass-radius relation 
$R/{\rm R_{\odot}}=\left( M/\msun \right)^{0.75}$ \citep{HA:04}, 
we could assign a physical dimension to the particles and investigate
the probability of stellar collisions and mergers.
In detail, we defined the minimum impact parameter 
for a collision as 2$R$, and the two stars were assumed to coalesce 
when their relative velocity upon collision was lower than the escape 
velocity from their surface.
The  binary separations adopted in this work were close to the extremes 
of the interval within which HVSs can be produced: 
small semimajor-axes, $a_0 \lesssim 0.05\au$, result in contact binaries,
while for  $a_0 > 0.2\au$ few stars would be ejected with velocities 
sufficient to escape the Galaxy \citep{Al:05}.

In order to  treat also the case of unequal-mass binaries,  
we extended the work of Paper I to include an additional set of $\sim 2000$
integrations of binaries with component masses 
$M_{2}= 1-3\msun$ and $M_{1}= 6\msun$.
Initial conditions and results of these new runs are summarized in Appendix A.
In the following, for the case of unequal mass binaries,
we distinguish between the primary and secondary stars'  quantities  using the 
subscripts  $1$ and  $2$ respectively.

The orbital initial conditions that we adopt in the SPH simulations 
correspond to  binaries that enter well within their tidal disruption 
radius $r_{\rm bt}$,
approximated  by the expression \citep{ML:05}:
\begin{equation}\label{rbt}
r_{\rm bt} \sim \left( \frac{M_{\bullet}}{M_{\rm b}} \right)^{1/3}a_0,
\end{equation}
with $M_{\rm b}=M_1+M_2$ the total mass of the binary.
Therefore, the binaries in all of our simulations are 
strongly perturbed at the first periapse passage. 
Moreover, we focus on cases where the tidal perturbations 
from the SMBH on the single stars are expected to be significant. 
This corresponds to binaries with periapsides that lie close to
the tidal disruption radius  of a single star, or
\begin{equation}\label{rt}
r_{t} \sim \left( \frac{M_{\bullet}}{M} \right)^{1/3}R.
\end{equation}

We selected three types of initial condition that  can be 
classified according to  the final outcome of the  SPH simulations.

\noindent $1$. {\it Stellar collision} (without merger).
In some cases, gravitational perturbations from the SMBH can 
lead to a physical  collision between the two stars.
If the relative velocity at collision is sufficiently high, 
the stars can survive the interaction and avoid a merger. 
Soon after the collision, one star can be ejected at high velocity.

\noindent $2$. {\it Stellar merger}. 
If the two stars collide with a relative velocity at impact
smaller than approximately  the escape velocity from their surfaces, 
coalescence occurs, resulting in the formation of a new, 
more massive star. 
The merger remnant will remain bound to the SMBH unless 
extreme mass loss occurs during coalescence.

\noindent $3$. {\it (Clean) ejection of a HVS}. 
The tidal breakup of the binary by the SMBH results in the ejection
of a star at very high velocity. 
The former companion of the ejected star loses energy in the process 
and is deposited onto a tight orbit around the SMBH. 
Here, ``clean'' means that the member stars  of the binary do not 
collide with each other during the process of ejection.

For the sake of simplicity, in all the SPH simulations we chose the
initial apoapsis of the external binary orbit to be
$d=2000\au$  $\approx 0.01$ pc 
(except for case H8 which has $d=1700\au$; see Table 1).
However, starting the SPH simulations from this distance 
would greatly increase the required  computational  time.
The stars were therefore initially placed 
at a point of the orbit corresponding to a much smaller distance
from the SMBH: $r_0 = 5r_{\rm bt}$.  
This choice for $r_0$ was motivated by the fact that at these distances, 
the tidal forces from the SMBH are still too weak to significantly
influence the internal  structure of the stars. 
In addition, since $r_0$  is considerably larger than  $r_{\rm bt}$,  
at the initial time the internal binary eccentricity is still close to zero  
($e \lesssim 10^{-2}$).
As in Paper I, the plane of the binary's orbit about  the SMBH 
is the $x-z$  plane.

\begin{table*}\scriptsize
\caption{ \label{t1}}Summary of the SPH simulations.
\begin{center}
\begin{tabular}{lllllllll}
 \hline
 ${\rm Run}$  & $M_1(\msun)$ & $M_2(\msun)$ &  $a_0$(au)  &  $r_{\rm per}$(au)  & $\lambda_1$ ($\lambda_2$) & $\zeta_1(\zeta_2)$&SPH & $N$-body    \\
 \hline
 C1      & 3 &  3  & 0.05   &  1.50  &  1.36  &0.668& Collision+HVS &   Collision+HVS \\
 C2      & 3 &  3  & 0.2     &  2.67  &  2.43  &1.19& Collision+HVS &   Collision+HVS \\
 C3      & 6 &  6  & 0.05   &  1.50  &  1.07  &0.526&  Collision+HVS &   Merger              \\
 C4      & 6 &  6  & 0.2   &  2.04  &  1.46  &0.716&   Collision+HVS &   Collision+HVS  \\
 C5      & 6 &  3  & 0.1     &  5.61  &  4.01 (5.09) & 1.96 (2.49) &   Collision+HVS (primary) &   Collision+HVS (primary) \\

 \hline
 M1      & 3 &  3  & 0.05   &  5.05  &  4.59    &2.25& Merger    &   Merger \\
 M2      & 3 &  3  & 0.05   &  1.50  &  1.36 & 0.668&   Merger  &   Merger  \\
 M3      & 3 &  3  & 0.05   &  4.18  &  3.79 &  1.86 & Merger  &   Collision+HVS         \\
 M4      & 3 &  3  & 0.2     &  4.18  &  3.79  &  1.86        &  Merger &   Merger \\
 M5      & 3 &  3  & 0.2     &  38.2  &  34.7   & 17.0 &  Merger  &   Merger \\
 M6      & 6 &  6  & 0.05   & 2.67   &  1.91 &  0.935  &Merger  &   Collision+HVS  \\
 M7      & 6 &  6  & 0.05   & 2.67   &  1.91 &  0.935  & Merger  &   Merger \\	
 M8      & 6 &  6  & 0.05   & 4.18   &  2.99 & 1.46&  Merger  &   Merger  \\
 M9      & 6 &  6  & 0.05   & 5.05   &  3.62 & 1.77&  Merger  &   Merger \\	
 M10      & 6 &  6  & 0.2   & 2.04   &  1.46 & 0.716&Merger  &   Merger  \\
 M11     & 6 &  6  & 0.2    &  20.4  &  14.6 & 7.13 & Merger  &   Merger \\
 M12     & 6 &  3  & 0.1    & 8.08   & 5.78  (7.34)& 2.83 (3.59) & Merger & Merger   \\	
 M13     & 6 &  1  & 0.1    & 3.48  & 2.49 (5.28)  &    1.22 (2.59)       & Merger & Merger  \\
  \hline 
 H1      & 3 &  3  & 0.05   &  1.50   &  1.36   & 0.668  &HVS & HVS  \\
 H2      & 3 &  3  & 0.05   &  7.07   &  6.42   & 3.14 &HVS & HVS  \\
 H3      & 3 &  3  & 0.2     &  2.04   &   1.86  & 0.909 & HVS & HVS  \\
 H4      & 6 &  6  & 0.05   & 1.50   &  1.07    & 0.526 &HVS & HVS  \\
 H5      & 6 &  6  & 0.2     &  2.04  &    1.46  &  0.716 & HVS & HVS  \\
 H6      & 6 &  6  & 0.2     & 0.60   &    0.429    &  0.210 & HVS & HVS   \\	
 H7     & 6 &  1  & 0.1     & 1.44    &    1.03   (2.19)   &   0.503 (1.07) & HVS (secondary)& HVS (secondary)  \\
 H8      & 6 &  1  & 0.1     & 0.795   &    0.569   (1.21)   &    0.278   (0.592) & HVS (secondary) & HVS  (secondary) \\	
 \hline \\ \\
\end{tabular}
\end{center}
\end{table*}

\subsection{SPH numerical techniques and initial conditions}

SPH is a Lagrangian method in which the fluid is represented by a finite number of
fluid elements or ``particles.''  Associated with each particle $i$
are, for example, its position ${\bf r}_i$, velocity ${\bf v}_i$, and
mass $m_i$.  Each particle also carries a purely numerical smoothing
length $h_i$ that determines the local spatial resolution and is used
in the calculation of fluid properties such as acceleration and
density.  For a recent review of SPH, see \citet{2009NewAR..53...78R}.
The SPH code used in this work is presented in
\citet{2010MNRAS.402..105G}, with the augmentation that the analytic
solution to the Kepler two-body problem can be used to advance a star
bound to the SMBH through those portions of the orbit when hydrodynamic
effects are negligible (see \S\ref{BP}).
The equation of state is ideal gas plus radiation pressure and radiative cooling and heating is neglected.
To calculate the gravitational accelerations and potentials,
we use direct summation on NVIDIA graphics
cards, softening with the usual SPH kernel as in Hernquist
\& Katz (1989). Thus, gravity is softened only in interactions between neighbors and it is
 softened by exactly the SPH kernel of each particle.  As the smoothing
length of a particle changes, so does its softening.
The use of such a softening with
finite extent (as opposed, for example, to Plummer softening)
increases the accuracy and stability of SPH
models, consistent with the studies of Athanassoula et
al. (2000) and Dehnen (2001).
 
In our simulations, the SMBH is a compact object particle that
interacts gravitationally, but not hydrodynamically, with the rest of
the system.  The gravity of the SMBH is softened according to a
density profile defined by the standard SPH cubic spline kernel with a
constant smoothing length $h_{\bullet}=20{\rm R_\odot}$.  This 
approach has the advantage that the treatment of gravity is unsoftened for
separations $r>2h_{\bullet}$.
We note that $h_{\bullet}$ is small compared to the periapsis
separation $r_{\rm per}$ in all cases, so our code
is able to follow bound stars around the black hole for many orbits without
introducing spurious secular effects from gravitational softening.
The SMBH is allowed to move in response to
gravitational pulls.  However, because the $4\times 10^6 \msun$ SMBH
is much more massive than any of the binaries being considered, the
SMBH always remains very near the center of mass of the system, which
we take to be the origin.

Before simulating the interaction of a binary with the SMBH, we must
first prepare an SPH model for each binary component in isolation.  To
compute stellar structure and
composition profiles, we use the TWIN stellar
evolution code \citep{1971MNRAS.151..351E, 2008A&A...488.1017G,
  2008PHD_GLEBBEEK} from the MUSE software environment
\citep{2009NewA...14..369P}. We evolve
main-sequence stars with initial helium abundance $Y=0.28$ and
metallicity $Z=0.02$.  The $3 \msun$ star is evolved to an age of
50 Myr, yielding the same $2.15 {\rm R_\odot}$ (0.01 au) radius as in the
corresponding models of Paper I. The 1 and $6 \msun$
stars are each evolved to 18.2 Myr, yielding stellar radii of $0.891$
and $3.44{\rm R_\odot}$ respectively.  Figure \ref{fig:threeparents} shows
the resulting composition profiles for our 1,
3, and 6 $\msun$ stars, colored red, green, and blue respectively.

Initially, we place the SPH particles on a hexagonal close packed
lattice, with particles extending out to a distance only a few
smoothing lengths less than the full stellar radius. After the initial
particle parameters have been assigned according to the desired
profiles from TWIN, we allow the SPH fluid to evolve into hydrostatic
equilibrium.  During the relaxation calculation, the drag force
we include is the normal artificial viscosity but in the acceleration
equation only.

Figure \ref{fig:vlnb} shows energies versus time both during and after
the relaxation process.  From the internal energy $U$ and potential
energy $W$ curves, it is apparent that the star oscillates on a
hydrodynamical timescale, specifically with
a fundamental period of about 0.08 days.  During relaxation, these
oscillations are damped by the drag force, which does negative work on
the system and decreases the total energy $E$ toward that of a minimum energy
equilibrium state.  At a time of 1.84
days ($=100 G^{-1/2}\msun^{-1/2}\rsun^{3/2}$), the drag force is
removed and the star is allowed, as a test of stability, to evolve
dynamically in isolation.  During this dynamical evolution, the
internal energy $U$ and gravitational energy $W$ each remain nearly
constant.  By $t=9$ days, the kinetic energy $T$ has diminished to
nearly 10 orders of magnitude less than the total energy $E$ in
magnitude, corresponding to an exceedingly small amount of noise in an
otherwise static model.  
The overall level of energy conservation is excellent: extrapolating
forward the drift in
total energy $E$, which is linear in time, we find it would take 
about $2.4\times10^5$ days (660 years or $3\times 10^6$ oscillation periods) of
hydrodynamical evolution to reach
a 1\% error in total energy.

Our approach allows the parent stars to be modelled very accurately.
As an example, Figure \ref{fig:plnc100_m6}
plots both desired profiles and SPH particle data for the $6 \msun$ star.
The structure and composition profiles of the SPH model closely follow
the desired TWIN profiles.
Our relaxed models remain static and
stable when left to evolve dynamically in isolation: indeed, the
particle data shown in the lower panels of Figure \ref{fig:plnc100_m6}
are nearly identical to those in the upper panels, demonstrating that
there are no significant changes in the model even after more than 4000
days (or equivalently $5\times 10^4$ oscillation periods) of hydrodynamical evolution.

The binaries in our simulations are then created simply by shifting
two stellar models, each taken from the end of a relaxation calculation, to the
appropriate initial position and velocity provided by the N-body code.
We begin with binary components irrotational in the
inertial frame, which allows us to more easily study any rotation
imparted during the subsequent interaction.

Unless otherwise noted, our simulations employ $N\approx 4\times10^4$
SPH particles: such a particle number provides an appropriate balance
between resolution and the need sometimes to follow the hydrodynamics for
time intervals exceeding $10^5$ dynamical timescales (corresponding to hundreds of orbits around the SMBH).

\subsection{Timescale considerations and orbital advancement}

Because of shock heating in collisions and mergers, in addition to
tidal heating during the periapse passage, the bound stars are out of
thermal equilibrium and larger than a normal main-sequence
star of the same mass.  The global thermal readjustment of the bound
stars proceeds on a thermal timescale $t_{\rm thermal}\approx U/L$, where
$U$ is the total internal energy in the star and $L$ is its
luminosity.  The SPH simulations confirm that the internal energy $U$ of
the bound star after one periapsis passage is comparable to the total
internal energy of the star(s) from which the bound star came and
typically $U\approx (2-7)\times 10^{49}$erg, with the larger values
generally corresponding to more massive stars.  For weak collisions
and clean ejections of HVSs, the luminosity of the bound star will be
comparable to the value it had in the initial binary: the thermal
timescale in such cases is then roughly $10^5$ to $10^7$ years.
Guided by calculations of blue stragglers \citep{1997ApJ...487..290S},
we estimate that the luminosity of a bound star produced by a merger
or strong collision may be up to $\sim 100$ times larger than that of
a main-sequence star of the same mass.  Thus, the luminosity of our
most massive merger products could be briefly as large as $\sim10^6
{\rm L_\odot} \approx 4\times 10^{39}{\rm erg~ s}^{-1}$, so that the global
thermal timescale $t_{\rm thermal}\gtrsim 600$ years (although the local
thermal timescale in the outer layers of the star could be less).
We conclude that thermal adjustment over an orbital period is small and often
completely negligible, and we therefore do not attempt to model the thermal
relaxation here.

Although the orbital period is small compared to the thermal
timescale, it is large compared to the hydrodynamical timescale, which
is about an hour.  Following the full hydrodynamics of a multiple
orbit encounter would therefore not be practical.  What we do instead
is wait for the star(s) to move sufficiently far away from the black
hole and then advance any bound star around most of its Kepler two
body orbit.  At the same time, we remove from the simulation any HVS,
any ejecta, and any gas that has become bound to the SMBH.  As long as
the periapse passages are treated hydrodynamically, our results are
not sensitive to precisely which portion of the orbit is treated in
the two body approximation.  In practice, we wait at least 8 days
after periapse, and at least 8 days after the merger or ionization
of a binary, before measuring the orbital elements and implementing
the two body analytic solution.  The orbital advancement is performed
such that the distance from the black hole to the bound star is
unchanged but that the objects are now approaching one another.  We
preserve the orientation of the orbit and spin of the bound star
during the orbital advancement.

 In order to test the reliability of the method, we run a portion of the first orbit for one of our simulations (C1 in Table 1) 
without any orbital advancement.  We found that, at about 20 days after we would have applied the advancement, 
their masses  had decreased by  an additional $0.002\msun$.
We note that even though we retain  too much mass,  this "extra" mass is far from the stars,
so it doesn't significantly participate in the hydrodynamics
and it will get ejected in the next passage. We conclude that neglecting hydrodynamics far from periapsis
is a reasonable approximation.

\section{Results}
The first important result of the SPH simulations performed in this work is that 
their qualitative outcome agrees very well with that of the N-body integrations devised in Paper I.
Among a total of 26 simulations, only in 3 cases did the N-body approach fail to  match the 
results of the hydrodynamic calculations.  

Table 1 reports  the chosen  initial conditions  as well as a qualitative  description of the final outcome 
of the SMBH-binary interaction in both SPH and $N$-body simulations.
The strength of the SMBH-stars interaction is parameterized by the dimensionless quantity: $\lambda= r_{\rm per}/r_{\rm t}$.
In general, stars on orbits meeting the condition $\lambda<1$ are tidally  disrupted  \citep{LC:86,  EK:89}.
However, even when  tidal disruption does not occur, we expect
that mass will be stripped from the outer regions of any star passing within its Roche  limit \citep{pac:71}.
In the table,  $\zeta$ gives the  Roche lobe radius (evaluated at periapsis) in units of the stellar radius:
\begin{equation}
\zeta=\frac{0.49q^{2/3}}{0.6q^{2/3}+{\rm ln}(1+q^{1/3})}\times
\frac{r_{\rm per}}{R}~~, \label{zeta}
\end{equation}
with $q=M/M_{\bullet}$ \citep{egg:83}.
Although this formula was derived under the assumption of circular orbit, it has been shown to
work reasonably well even for eccentric binaries,  if  used at periapsis \citep{RS:05}.
Note that because $q<<1$ in the present work, the approximate relation $\zeta \approx 0.49
\lambda$ exists between $\lambda$ and  $\zeta$.

The results of our SPH simulations are presented in what follows.
We first describe the product of one binary-SMBH interaction (Sections \ref{COLL}, \ref{MG} and \ref{HVS}), and then we successively follow 
the evolution of the  bound stars  as they perform several revolutions around the SMBH (Section \ref{BP}).

\begin{table*}\scriptsize
\caption{ \label{t2}} 
{Ejection velocity $v_{\rm ej}$  of the HVS, orbital semimajor-axis $a$ and eccentricity $e$
of the captured star, and distance of closest approach between the two stars ($r_{\rm 0}$) in  the SPH ($N$-body) calculations. 
The quantity  $\Delta M_{\rm b}/M_{\rm b}$  gives the fraction of mass 
lost from the binary, while $\Delta M_{\bullet}/M_{\rm b}$ is the fraction of the mass lost from the binary  that remains  bound to the SMBH.  
All quantities are evaluated after the first periapsis passage, once the stars have retreated far from the SMBH.}
\begin{center} 
\begin{tabular}{llllllll}
 \hline
 ${\rm Run}$  & $v_{\rm ej}$ &      a                      &  e   &   $r_{0}/(R_1+R_2)$& $\Delta M_{\rm b}/M_{\rm b}$              &  $\Delta M_{\bullet}/M_{\rm b}$ \\
                         &     (km/s)       &     (au)                  &           &       &             	         			    &                                            &      \\
 \hline
 C1      &  4608(5681)&  124(101)   &0.988(0.985)  &  0.23(0.32) &$-1.33\times 10^{-2}$          &  $7.19\times 10^{-3}$    \\
 C2      &  3878(3964)&  159(157)    & 0.983(0.983) & 0.49(0.50)& $-2.80\times 10^{-3}$          &  $2.06\times 10^{-3}$      \\
 C3      &  3335(4840)&  190(117)    & 0.991(0.987) & 0.38(0.45)        & $-1.24\times 10^{-2}$         &  $6.73\times 10^{-3}$    \\
 C4      &  1467(1554)&  380(377)    & 0.995(0.995) & 0.67(0.87)& $-1.29\times 10^{-3}$         &  $7.08\times 10^{-4}$  \\
 C5      &  1764(2367)&  212(163)    &  0.974(0.965) &0.42(0.61) &$-2.16\times 10^{-3}$          &  $1.53\times 10^{-3}$   \\
\hline
\end{tabular}
\end{center}
\end{table*}

\subsection{Stellar Collisions} \label{COLL}
\citet{GL:07}  noted that the tidal breakup of 
stellar binaries interacting with the SMBH can lead, under some circumstances, 
 to a physical collision between the two member stars.
In this section we investigate the cases in which  the two stars
collide with a relative impact speed large enough that they do not merge upon impact.  
Some results of the SPH simulations are shown in Table  \ref{t2}, where we also compare  the asymptotic 
ejection velocity of the HVSs ($v_{\rm ej}$), the semimajor-axis ($a$) and eccentricity ($e$) of the captured stars
with the same quantities obtained in the $N$-body simulations.

The agreement between the two methods is remarkably good.
However, the SPH simulations systematically 
produce  smaller values of  $v_{\rm ej}$ and larger $a$.
This is a consequence of the efficient 
energy and angular momentum transfer  occurring  between the stars:
the ejected star slows down and the captured star  gains speed during the collision.

  From  Table  \ref{t2}, it is clear that the mass-loss from the binary is smaller than  a few percent of the total mass, and always below 
$\sim 0.1 \msun$.  
This is consistent with previous simulations of  stellar collisions  that often found a small
fractional  mass-loss \citep{BH:87,BH:92,FB:05}.  
Our   calculations indicate that, after a collision, the mass ejected
from the SMBH-stars system is comparable to, although smaller than,
the  mass that is ejected from the binary but remains bound to the
SMBH; this mass-loss has a small but measurable impact on the subsequent evolution  of the stars' orbits  as 
demonstrated by comparing the SPH and the $N$-body quantities in the table.
Debris will eventually settle into a torus-like structure about the SMBH, that will subsequently  evolve  due to viscosity, mass inflow, 
radiative cooling, and winds.

Spin-up is  expected to be one of the main signatures of either a (off-axis)  collision  \citep{AK:01} or  a tidal encounter with a massive black hole  \citep{EK:89}.
In our simulations the   close stellar encounter as well as the  SMBH tides at periapsis lead therefore to some degree of rotation in the stars with  angular frequency:
\begin{equation}\label{rotaz}
\Omega_{\rm tot} \approx    \sqrt{\Omega_*^2     +   \Omega_\bullet^2}  \approx  \sqrt{\frac{G(M_1+M_2)}{r_{\rm 0}^3}+ \frac{GM_{\bullet}}{r_{\rm per}^3}}~~, 
\end{equation}
where $  \Omega_*$  and $ \Omega_\bullet   $ are respectively  the angular velocity induced by the interaction with the companion star and that imparted by the 
SMBH tides; $r_{\rm 0}$ is the distance of closets approach  between the stars.
The ratio $  \Omega_*/\Omega_\bullet$  in the cases considered here is always grater  than 1 and varies from a maximum of $\sim 20$ (simulation C5)
to a minimum of $\sim 2$ (simulation C4).

\begin{figure}{~~}
\begin{center}
\includegraphics[width=3.in]{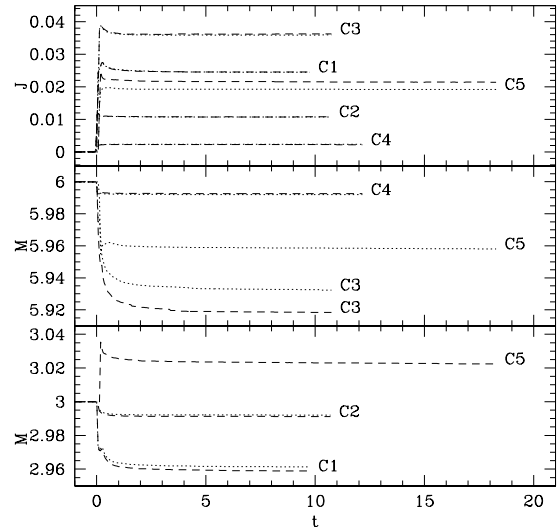} 
\caption{Temporal evolution of the  dimensionless spin parameter $J$ defined in equation (\ref{PP}) and the stellar masses $M$ in solar units at times near the first periapsis passage in simulations with stellar collisions.
Dashed curves correspond to the stars captured by the SMBH, while the dotted curves are for the ejected stars. 
The time coordinate is given in units of days and is shifted in order to have $t=0$ at the moment of the closest approach of the binary  with the SMBH.  The curves terminate at the time the orbit is advanced using the analytic two-body solution.
}
\label{rotC}
\end{center}
\end{figure}

Figure \ref{rotC}  plots the temporal evolution of the dimensionless spin parameter \citep{PE:71}:
\begin{equation}\label{PP}
J=\frac{L|E|^{1/2}}{GM^{5/2}}~~,
\end{equation}
where $L$ is the spin angular momentum of the star and $E$ its binding energy.
In all cases,   there is a sharp  increase in the stars' spins during the periapsis passage followed
by a second gradual decrease  toward the final relaxed (spinning) configuration.
Note that the captured stars have values of the final spin slightly larger than that  of  the ejected stars.
This finding is consistent with the results of paper I (but see \citet{Sari:2010} as well), where it is shown that 
the captured member is always the star with the smallest value of the closest approach distance to the SMBH implying, 
as we expect  from  equation (\ref{rotaz}),  a larger tidal torque at periapsis.

Figure \ref{rotC} also shows the temporal evolution of the stellar
masses near the first periapsis passage.  Each star typically loses
about 1\% of its mass, with captured stars (again, those that pass
closer to the SMBH) losing slightly more mass than their ejected
counterparts.  In the cases with an equal mass binary, both stars lose
a comparable amount of mass.  The captured star in simulation C5
actually gains mass that had been lost from its binary companion, as
discussed in more detail below.  In all cases, the stellar masses
stabilize to an essentially constant value by the time the orbital
advancement technique is implemented, which is where the curves
terminate.

\begin{figure*}{~~}
\begin{center}
\includegraphics[angle=270,width=6.5in]{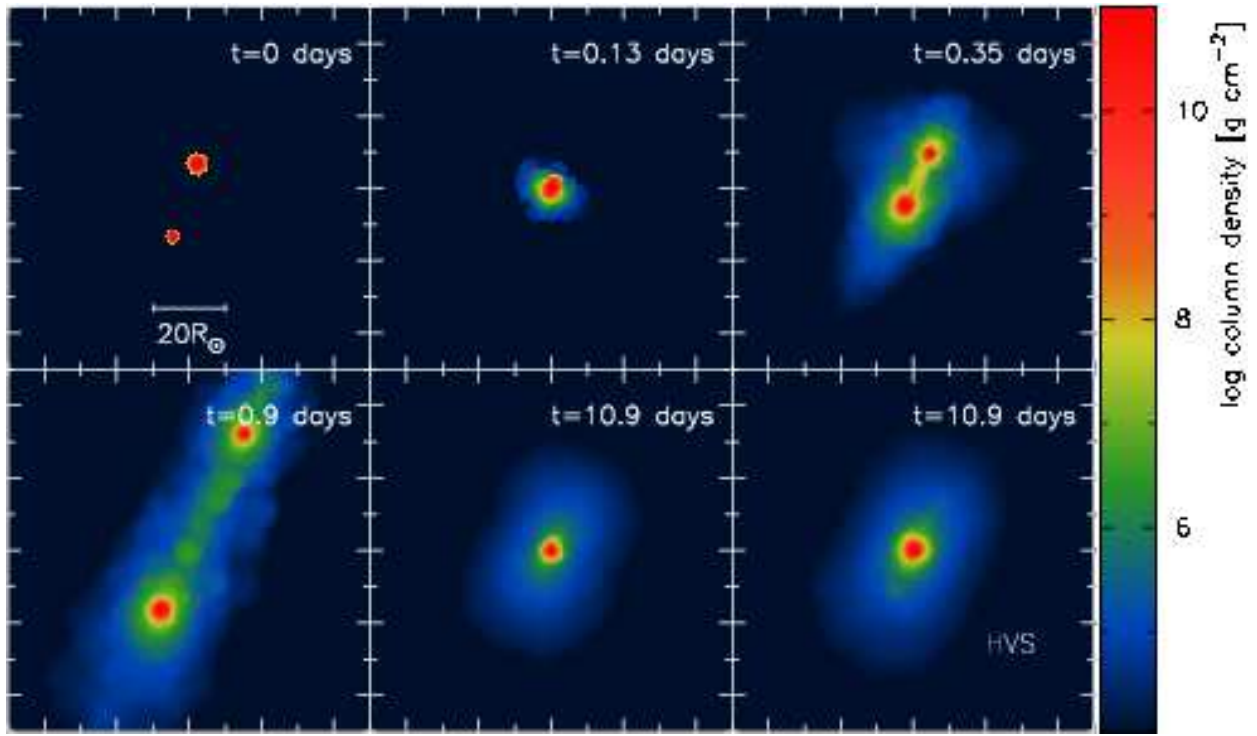} 
\caption{Column density plots for simulation C5 on the $X-Z$ plane. In this case the binary has an internal semimajor-axis $a_0=0.1$au
and its components have masses  $M_1= 6 \msun$ and  $M_2= 3\msun$.  
Time $t=0$ corresponds to the periapsis passage of the binary external orbit.
Between $t=0$ and $t=0.9$ days the panels are centered on the binary center of mass. The two bottom-right panels 
are centered on the center of mass of either the captured (left) or ejected (right) star. 
The black hole is outside the images.
At $t=0.13$ days the stars collide. Subsequently, the $6 \msun$ member is ejected at hypervelocity while
the secondary star remains bound to the SMBH. }
\label{coll5}
\end{center}
\end{figure*}

Figure \ref{coll5} presents column density snapshots for simulation C5.
The simulation models a stellar binary with $a_0=0.1$ au and with components of masses 
 $M_1= 6 \msun$ and  $M_2= 3\msun$ (see Table 1).  
The time indicated on the panels is shifted in order to have $t=0$ at the moment of the
closest approach of the binary with the central black hole.
Between $t=0$ and $t=0.9$ days the reference frame is the center of mass of the binary
while, in the lower-right panels, we switch to the frame in which the center of mass 
of either  the captured (left) or ejected (right) star is at the origin.

The first contact between the stars occurs $\sim 0.1$ days after the periapsis passage.
The interaction leads to an episode of mass transfer between the stars  (observed at $\sim 0.35$ days
in the figure).    The smaller star gains mass  ($\sim 0.02 \msun$) in
the collision, while the larger star loses $\sim 0.04 \msun$.
Only $6 \times 10^{-3} \msun$ becomes ejecta from the entire system (i.e., stars plus SMBH) after the first periapsis passage, while the remainder
of the gas lost from the binary  remains bound to the black hole. 
The mass loss from the binary  upon impact is therefore of order $\sim 10^{-2}\msun $.
Note that, in this simulation, the penetration factor $\lambda$ of both stars is large enough that the mass loss from the binary can be completely 
attributed to the stellar impact rather than to the tidal perturbations from the SMBH.
% In simulation C3, for instance, $\lambda \sim r_{\rm t}$ (see table 1) and the much higher mass loss
% ($\sim 0.1 \msun$) is mainly induced by the interaction with the SMBH.
 
\begin{figure*}
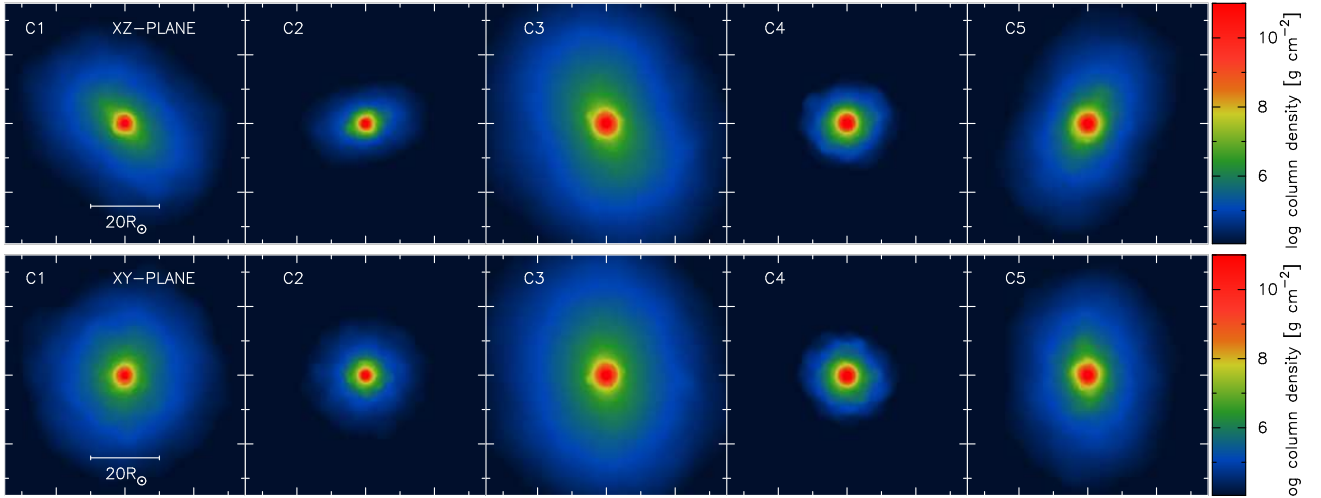
{~~}
\begin{center}
\includegraphics[angle=270,width=6.8in]{antonini_fig7a.eps} 
\includegraphics[angle=270,width=6.8in]{antonini_fig7b.eps} 
\caption{Stars ejected in our simulations after a collision with the companion star.
The stars show, typically, an oblate envelope surrounding a high density 
spherical nucleus.  Only in simulation C4, where the impact is  more ``grazing,''
 the collisional   product is spherically symmetric  even in its outermost envelope.
}
\label{CHVS}
\end{center}
\end{figure*}

The two bottom-right panels give column density plots of the captured (left) and ejected (right) stars, both showing 
a low-density, oblate  envelope surrounding  a compact spherical nucleus with a central density almost 
unaltered with respect  to that of the parent stars.
This particular configuration is common  to almost all the other collisional products as shown  in Figure \ref{CHVS} which
gives column density plots of the ejected stars.  
 Rotation as well as asymmetric shape  
 can have a fundamental role in the future evolution of the stars and their observable 
 characteristics;  a star 
 with a rapidly rotating nucleus can have its main-sequence life-time considerably extended with respect to their
 non-rotating counterpart \citep{CL:94}. 
Only in run C4 the HVS ejected after the collision 
is slowly spinning  and spherical even in its outermost envelope.  In this simulation,  the stars experience  a  more grazing collision which
leads to some envelope-ejection  but  leaves  the stars' structure essentially unchanged.

\begin{table*}\scriptsize
\caption{ \label{t3}} 
{Same as Table  \ref{t2} but for stellar mergers after the first periapsis passage. Here the quantities in parentheses refer to the 
initial  orbit of the  binary center of mass.}
\begin{center} 
\begin{tabular}{llllllll}
 \hline
 ${\rm Run}$  &      a                      &  e   & $\Delta M_{\rm b}/M_{\rm b}$              &  $\Delta M_{\bullet}/M_{\rm b}$ & $J$ \\
                        &     (au)                  &                  &             	         			    &                                            &\\
 \hline
 M1      &   1060(1000)   &0.995(0.995)   &$-2.57 \times 10^{-2}$          &  $2.57 \times 10^{-2}$    & 0.153   \\
 M2      &   984(1000)   & 0.999(0.999)  & $-1.61 \times 10^{-2}$          &  $1.26  \times 10^{-2}$  &  0.167   \\
 M3      &   997(1000)   & 0.996(0.996)& $-5.07 \times 10^{-2}$         &  $2.71\times 10^{-2}$       &  0.103   \\
 M4      &   1010(1000)    & 0.996(0.996) & $-4.26\times 10^{-2}$         &  $4.24 \times 10^{-2}$      &  0.259 \\
 M5     &   1020(1020)    &  0.958(0.963)  &$-2.16\times 10^{-2}$          &  $1.86\times 10^{-2}$      &  0.222     \\
 M6     &  996(1000)   &0.997(0.997)  &$-5.60\times 10^{-2}$          &  $ 2.93\times 10^{-2}$        &  0.0666          \\
 M7    &  1180(1000)    & 0.997(0.997) & $-2.50\times 10^{-2}$          &  $2.37\times 10^{-2}$          &  0.242 \\
 M8   &   1030(1000)    & 0.996(0.996)        & $-4.00\times 10^{-2}$         &  $2.06\times 10^{-2}$   &  0.249 \\
 M9      & 1000(1000)    & 0.995(0.995) & $-2.93\times 10^{-2}$         &  $2.65\times 10^{-2}$       &  0.238  \\
 M10     & 1020(1000)    &  0.998(0.998) &$-6.09\times 10^{-2}$          &  $5.99\times 10^{-2}$       & 0.108\\
 M11     &  1010(1010)   &0.980(0.982)  &$-4.38\times 10^{-2}$          &  $2.51\times 10^{-2}$      &  0.131\\
 M12    &  1020(1010)    & 0.992(0.991) & $-1.80\times 10^{-2}$          &  $1.23\times 10^{-2}$   &   0.181\\
 M13   &    1060(1000)    & 0.997(0.994)         & $-2.20\times 10^{-2}$         &  $1.63\times 10^{-2}$&  0.0766 \\
\hline
\end{tabular}
\end{center}
\end{table*} 

\subsection{Mergers}\label{MG}
Stellar collisions due to either binary evolution or dynamical
interactions are thought to be the main formation channel 
of blue stragglers in star clusters  \citep{CJ:84,Leonard:89,MAT:90}.
Similar processes have been proposed in the past to explain the puzzling presence of the  young massive 
stars observed at galactocentric  distances of few mpcs,  where star formation is thought 
to be strongly inhibited by the SMBH tides \citep{GE:03,EI:05}. 
In  paper I, we showed that gravitational encounters 
involving stellar binaries and the SMBH lead, for a wide range of orbital parameters, 
to a stellar collision and that among the collisional products, stellar coalescence 
occurs in more than $80 \%$ of the cases.         
In this section we study this latter outcome and investigate the properties of the resulting stars to clarify 
whether they would be expected to posses features commonly associated with the S-star population.

Table \ref{t3} gives the orbital parameters (i.e., eccentricity and semimajor-axis)
of the merger products in our simulations  as well as the mass ejected from the binary and the
fraction of mass captured by the SMBH after the first periapsis passage.
The table shows that the merger remnants lie on  a orbit very close to the
initial orbit of the center of mass of the binary around the black
hole, implying only a small effect of 
the mass-loss on the dynamical evolution of the stars.  
The mass ejected from the binary after the first periapsis passage is 
typically larger than that found in collisions that do not end up with a merger  (see Table \ref{t2}) and
is of order   $\sim 10^{-2}$ times the initial mass of the binary.
In many cases most of the mass ejected from the binary during the merger remained  bound to the SMBH.
This is a quite different situation respect to that found in Section  \ref{COLL}, where approximately  
half of the mass ejected from the binary remained  unbound to the black hole;
in these previous runs  one of the two stars is always
found on an escaping trajectory and, consequently,  the debris associated with such  a star
will also tend to escape the SMBH.
The last column in the table gives the  dimensionless spin parameter defined in equation  (\ref{PP}) of the final  merger products 
that show very large spins, some of them close to the ``break-up'' value (i.e., $J=1$). 

In principle it is possible that, as a consequence of  the mass loss occurring near periapsis, the resulting merger product  gains orbital
energy and escapes the SMBH (see for instance Faber et al.\ [2005]). 
Although we do not exclude this outcome for a different set of initial conditions, 
in our simulations this mechanism does not produce HVSs, and in all cases the merger remnant is still on a bound orbit around the SMBH.
Another more important consideration is that, subsequent to the merger, the tidal heating results in some degree of expansion
and a weakly   bound configuration for the merger product.  Because  the tidal radius of the newly formed star is
much larger than that of its progenitors,  the star  will successively
lose more mass with each periapsis passage and eventually be torn apart 
by the SMBH tides. And in fact, as it is shown  below, this is the final outcome of some of our simulations.

\begin{figure*}{~~}
\begin{center}
\includegraphics[angle=270,width=6.in]{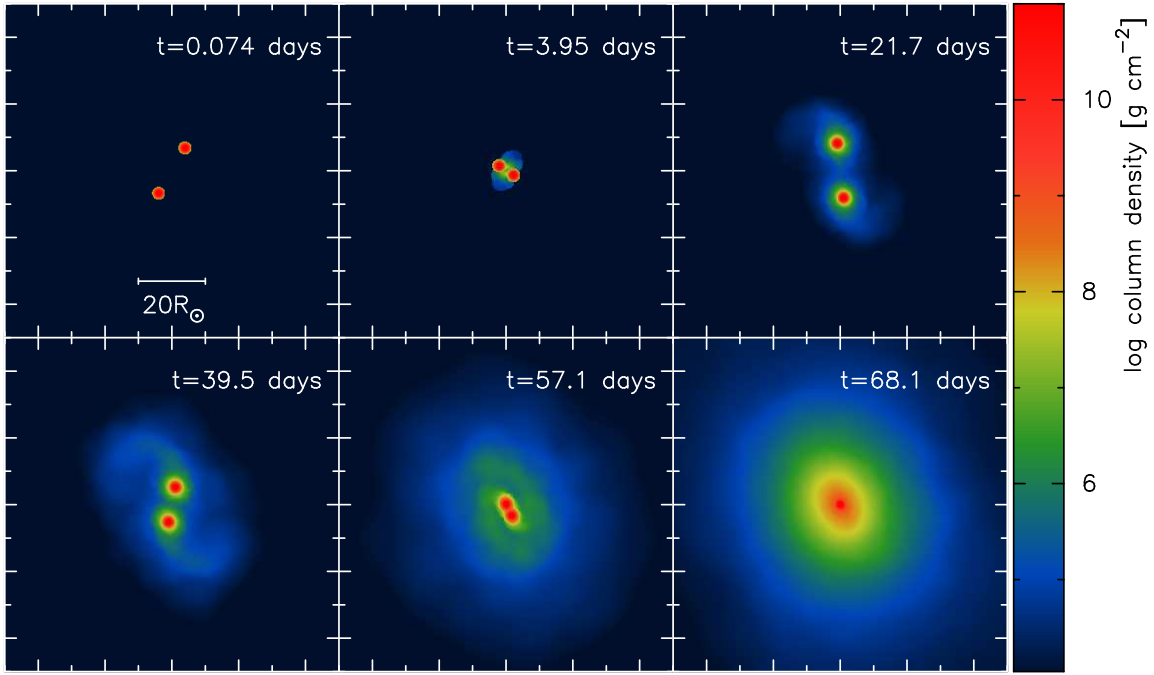} 
\caption{Column density plots for simulation M4 on the $X-Z$ plane. The binary has an internal semimajor-axis $a_0=0.2$au
and its components have masses  $M=3 \msun$.  
Time $t=0$ corresponds to the periapsis passage of the binary external orbit.}
\label{merg5}
\end{center}
\end{figure*}

An example of merger is displayed in Figure \ref{merg5}, which involves a binary with $a_0=0.2$ au and equal-mass components of masses $M=3 \msun$ (run M4).
In this run the stars collide for the first time after $\sim 4$ days from the time corresponding to the  periapsis passage;
the internal periapsis separation at the first contact is $r_{0}/(2 R)=0.86$.
After the first periapsis passage, as  consequence of the SMBH perturbation, the binary star becomes very eccentric. 
As the stars move through the circum-binary envelope
formed during the previous encounters, the orbit gradually circularizes and shrinks. By $t\approx 60$ days, after approximately $30$ collisions,   the two stellar nuclei  merge.
The final product has an oblate shape which is a common characteristic of all the merger remnants formed
in our simulations.

\begin{figure*}{~~}
\begin{center}
\includegraphics[angle=270,width=6.in]{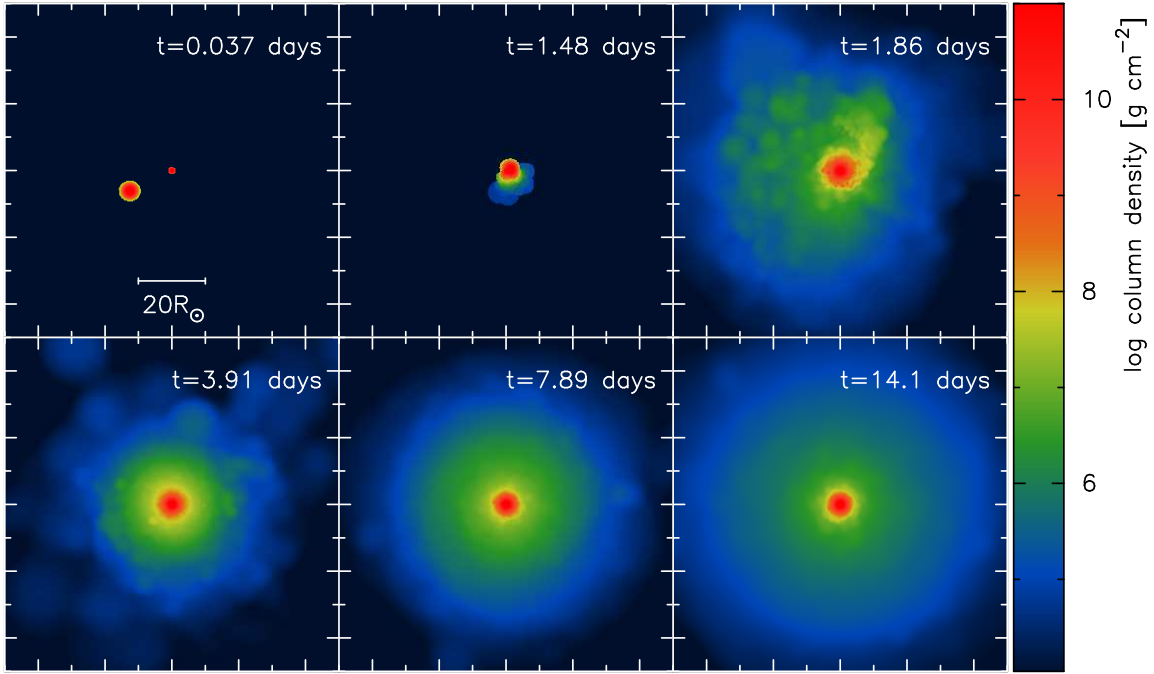} 
\caption{Column density plots for simulation M13 on the $X-Y$ plane. The binary has an internal semimajor-axis $a_0=0.1$au
and its components have masses  $M_1=6 \msun$ and $M_2=1 \msun$. 
Time $t=0$ corresponds to the periapsis passage of the binary external orbit.}
\label{merg13}
\end{center}
\end{figure*}

As another example,  Figure \ref{merg13} displays column density plots of  simulation M13 where a merger occurs between a $6$  and a $1 \msun$ stars.
The stars collide  after $1.48$ days from the moment of the closest approach to the SMBH, and  subsequently merge in the 
following $\sim 1$ day. 
During the merger,  the high density core of the lower mass star rapidly sinks to the center of the  companion star.
The   tail-like feature observed at  $t=1.86$ days in the figure, is mostly material coming  from the 
the secondary star that loses part of its outermost  envelope while  sinking  to the center of the merger remnant.

\begin{figure}{~~}
\begin{center}
\includegraphics[angle=0,width=3.2in]{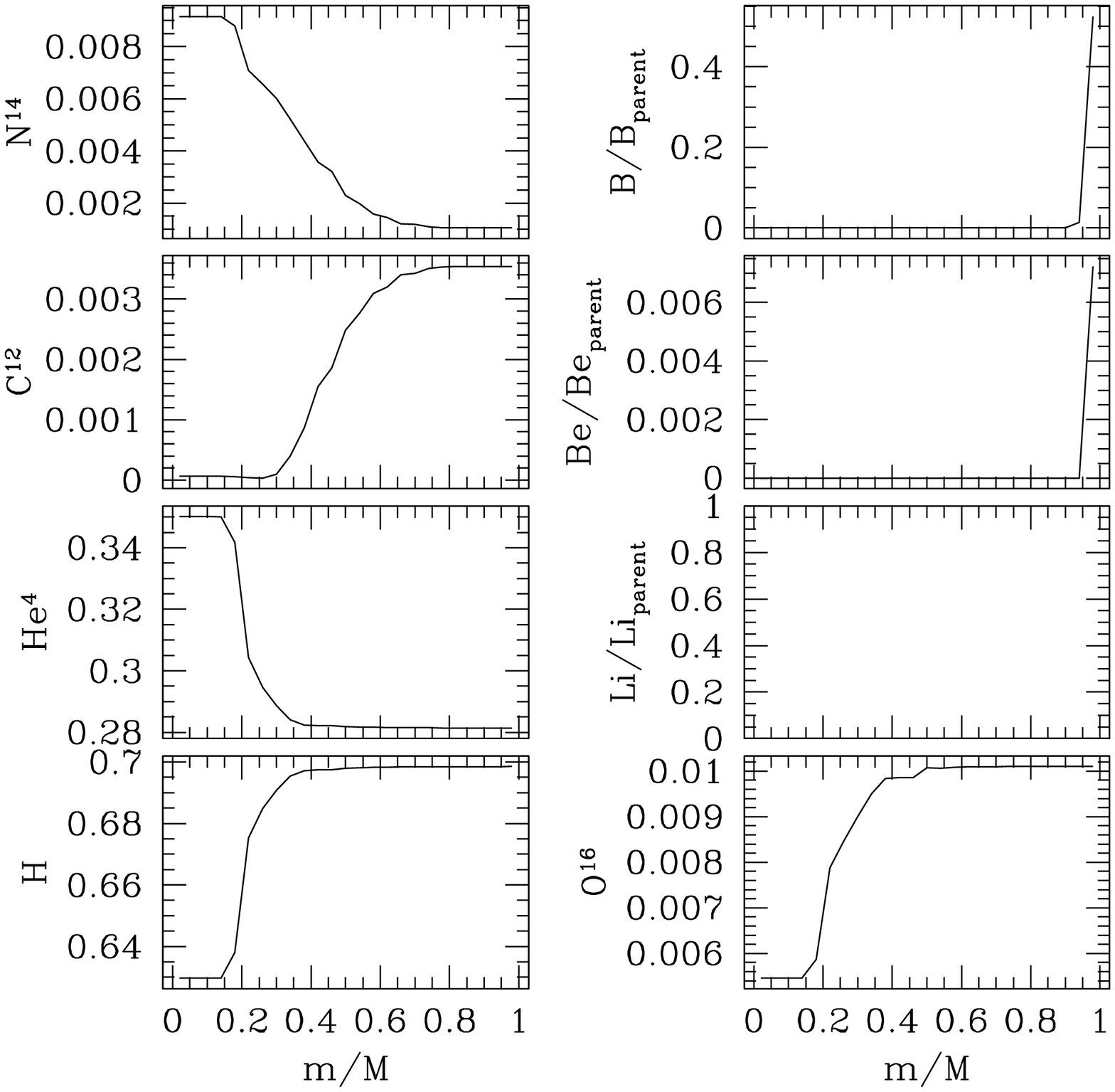} \\  \includegraphics[angle=0,width=3.2in]{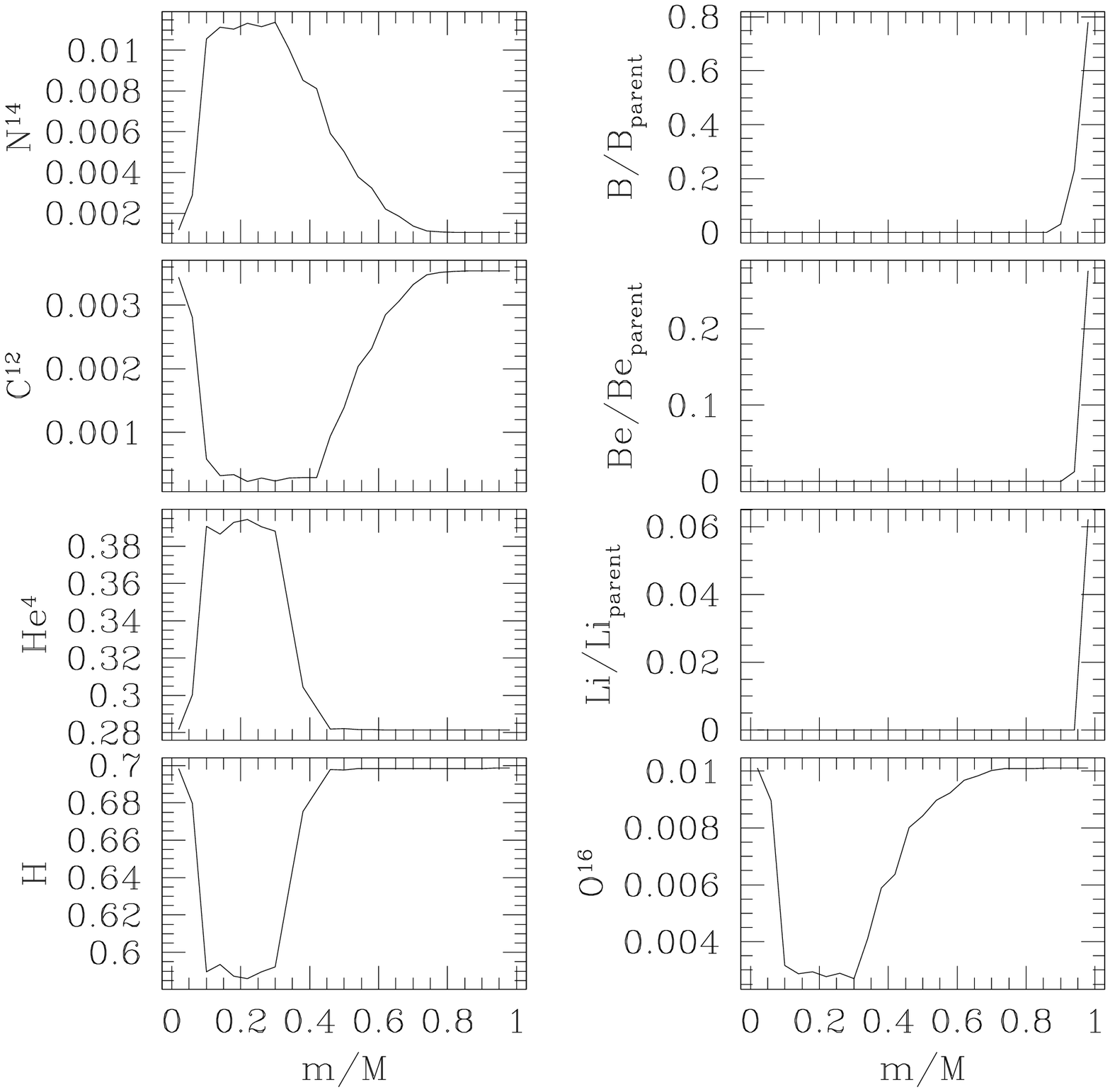} 
\caption{Upper panels: composition profile of the merger remnant
  formed after one periapsis passage in run M4 (see also Figure \ref{merg5}).
The merger product of two equal mass stars has a  composition profile very similar to that of its parent stars.
For a $\sim 6 \msun$ star, a final  He abundance in the core of $Y=0.35$ corresponds to an effective age of $\sim 2$ Myr.
Lower panels: composition profile of the merger remnant formed
  after one periapsis passage in run M13 (see also Figure \ref{merg13}).
In this case, the merger product  has a peculiar profile if compared to a ``normal'' star. Its core is strongly  
hydrogen-enriched as a consequence of the  low-He fluid
transported by the low mass star  along with it to the center.   }
\label{merg-comp}
\end{center}
\end{figure}

Figure \ref{merg-comp} shows chemical composition profiles of the
merger products for runs M4 and M13, after one periapsis passage.  
In simulation M4, the remnant has a mass of roughly $5.9 \msun$, its composition profile is very similar to that of the parent stars (see Figure \ref{fig:threeparents}).  
Based on how long it would take a normal star of that
mass to evolve to that central hydrogen abundance using the TWIN
stellar evolution code, we estimate that   a  ``normal''   $5.9 \msun$  star would reach a core helium abundance  $Y=0.35$ (assuming $Z=0.02$) after
$\sim 2$Myr.  By colliding two $50$ Myr old $3 \msun$ stars, we have effectively made a more massive ($M \sim 6 \msun$), younger (age $\sim 2$ Myr) star.
The merger remnant of run M13 shows a peculiar  composition profile when
compared to a normal star, but quite  normal for
merger products.  In M13 (and in M12 as well), the low mass star
drops to the center of the merger product
 bringing  its fresh hydrogen fuel along and significantly rejuvenating the core.
 As a result, the maximum He does not occur at the center of
the merger product.
The main reason of the negligible amount of hydrodynamic mixing is that the cores of the initial stars 
are very dense and difficult to break even in a head-on collision \citep{LO:95, LO:96}. 
However, we cannot exclude the possibility that other processes occurring on a thermal timescale 
(as opposed to the hydrodynamic timescale) can produce a significant degree of mixing 
in the stars as they evolve toward thermal equilibrium \citep{1997ApJ...487..290S}.

\begin{figure}{~~}
\begin{center}
\includegraphics[width=3.in]{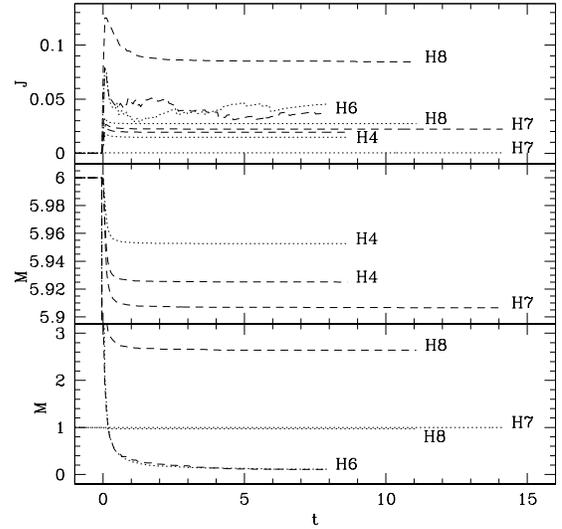} 
\caption{Like Fig.~\ref{rotC}, but for some of the simulations in which there is neither a 
stellar collision nor a merger.}
\label{rotH}
\end{center}
\end{figure}

Another interesting result, plotted in Figure \ref{merg-comp}, is that a large fraction
of the Lithium/Beryllium/Boron from the parent stars gets ejected,  indicating that 
a significant gap in the abundances of these elements in the S-star population  might  be observational
evidence of  rejuvenation through merger. Reduced atmospheric Lithium abundances   are for instance  observed in   blue  stragglers and
can also be a strong  indicator of mixing \citep{HM,PG:91}.

We finally  note that as the stars   keep orbiting the SMBH, their chemical profile and their spinning configuration 
will change in time and therefore the states displayed in Figures  \ref{merg5},  \ref{merg13} and  \ref{merg-comp} should be intended
as not permanent. 
The evolution will typically  lead  toward smaller spins and  a lower  
Lithium/Beryllium/Boron abundances in the stars.
We will come back to this point below.

%It is possible that the S-stars are drawn from the young population observed in the stellar 
%disk that extends inward to within $0.1~{\rm pc}$ of the SMBH, but while the S-stars are primarily 
%B dwarfs, the young stars in the disk are mainly luminous Wolf-Rayet and OB supergiants and giants  \citep{PA:01}.
%This limits the connection between the two stellar populations, as
%O/W-R stars are typically more massive and shorter-lived than B types.
%\citet{LBK} proposed that binaries scattered from the stellar disk on highly eccentric bound orbits around the SMBH
%would be disrupted, ejecting one member as a HVS while leaving the other member bound. 
%The inspiral of a cluster hosting an IMBH would generate stars on comparable orbits.
 
\subsection{Clean Ejection of Hypervelocity Stars}\label{HVS}
Even when the stars do not collide, tidal torque and mass-loss can occur if the periapsis distance of the binary center of mass
initially lies within the Roche limit of its member stars.
Similarly to what is done in the previous two  subsections, we analyze
here the first binary-SMBH interaction, while we discuss the following  evolution of the bound 
stars  in the next  subsection.
Some of the results of our SPH simulations, for which there is not a direct collision between the stars, are listed in Table
\ref{t3}.  
As expected, for $\zeta \gtrsim 1$ there is no mass-loss, and the stars maintain 
their  initial configuration essentially unaltered (runs H2 and H3).
Conspicuous mass-loss instead occurs  for runs H6 and H8 in which at least one of the stars crosses its tidal radius.
Interesting, although the usual condition for tidal disruption is well satisfied (i.e.,  $\lambda < 1$), in both runs  the stars   are not fully disrupted by the SMBH's tidal gravity at the 
first periapsis passage. 
In the table we also give the final values of the spin parameter $J$ that, in general, are found to be   a very small fraction of the 
 breakup value.  
Figure \ref{rotH} shows the temporal evolution of $J$ and the stellar masses for the cases of Table \ref{t3} that have the highest value of the final spin.
The interaction with the SMBH induces a strong rotation only in the
stars of runs H6 and H8.
We conclude that,   unless the stars penetrate deeply their tidal disruption radius, it seems  unlikely  that   the SMBH tides at periapsis alone
can produce a significant  spin-up  of the stars.

\begin{figure*}{~~}
\begin{center}
\includegraphics[angle=270,width=6.in]{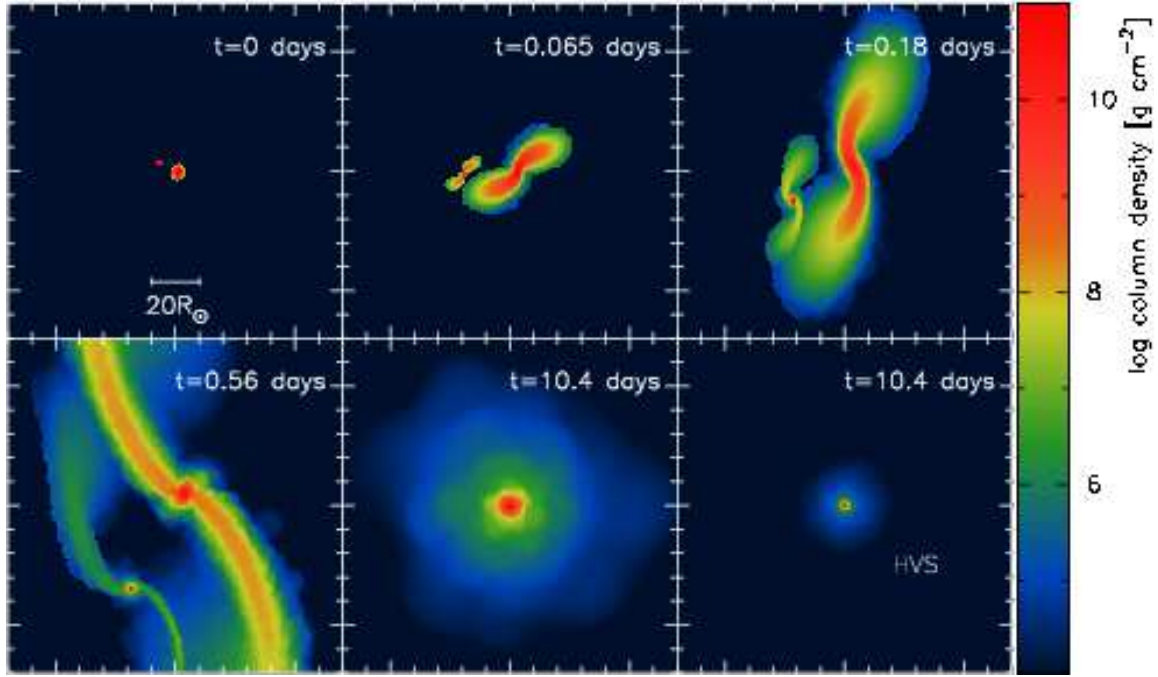} 
\caption{Column density plots for simulation H8 on the $X-Z$ plane. The binary has an internal semimajor-axis $a_0=0.1$ au
and its components have masses  $M_1=6 \msun$ and $M_2=1 \msun$. 
Time $t=0$ corresponds to the periapsis passage of the binary external orbit.
The first four panels are centered to center of mass of the binary, while in the
two bottom-right panels the origin is the center of mass of either the captured (left) or
ejected star (right).}
\label{h8}
\end{center}
\end{figure*}

As an example,  Figure \ref{h8} gives column density plots of run
H8. In this simulation, the primary and secondary stars have
masses $1 \msun$ and $6 \msun$  respectively. The periapsis of the external orbit ($\sim 0.8$au) 
is initially inside the tidal radius of the $6 \msun$   member but it is still outside the tidal radius of the $1 \msun$ star.
At periapsis, the stars are squeezed by the SMBH's tidal gravity. In the process the primary star losses 
a large fraction of its initial  mass ($\sim 3.4 \msun$), while the secondary loses only $\sim 0.03 \msun$.
After the interaction with the SMBH the binary is broken apart and the lightest member becomes an HVS.
Because of tidal heating during the periapsis passage,  the stars are perturbed from their thermal equilibrium 
state  and their radii are  somewhat enlarged with respect to a normal main-sequence star of the same mass.

\subsection{The Bound Population} \label{BP}
After the initial encounter between the binary and the SMBH, one star
remains in a bound orbit around the SMBH in all of our simulations.
Like the orbit of the initial binary about the SMBH,
the orbit of such a bound star is highly eccentric:
$0.96<e<1$.  In those cases in which the bound star is a merger product
(runs M1 through M13), the semimajor-axis $a$ very nearly equals the
semimajor-axis of the initial binary about the SMBH: $a\approx
1000{\rm au}$, corresponding to an orbital period of about $16$ years.
We note that these orbital periods are comparable to those of the
S-Stars in the GC.
In those cases in which a HVS star is ejected (runs C1 through C5 and
runs H1 through H8), the ejection energy comes at the expense of the
orbital energy of the bound star, which consequently has a somewhat
smaller semimajor-axis: $100{\rm au}\lesssim a\lesssim 700{\rm au}$,
corresponding to orbital periods of 0.5 to 9 years.

\subsubsection{Tidal Stripping}

The most significant hydrodynamic effects occur near the periapsis,
where induced collisions and mergers are most likely to occur and
where tidal stripping is at its greatest.

\begin{figure}
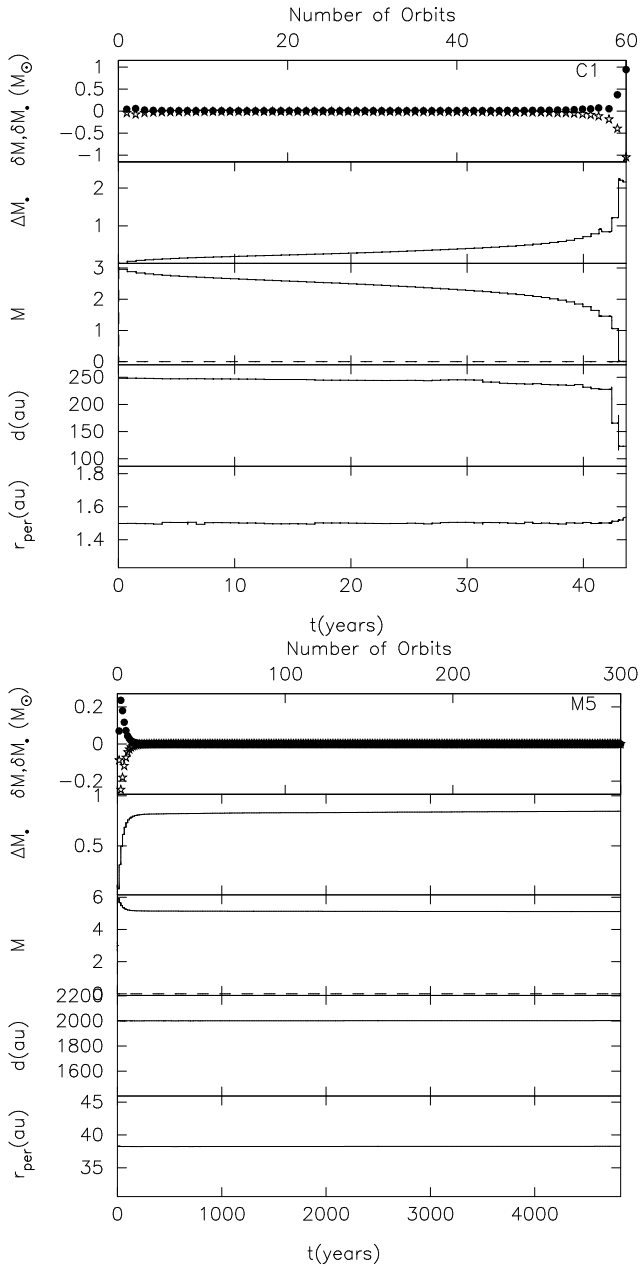

  \begin{center}
     \includegraphics[angle=270,width=84mm]{antonini_fig13a.eps}
     \includegraphics[angle=270,width=84mm]{antonini_fig13b.eps}	
  \end{center}
  \caption{Evolution versus time (and number of orbits) in the runs C1 and M5.
From the top to the bottom panel:
 mass $\delta M_\bullet$ gained per orbit by the SMBH
(filled circles) and mass $\delta M$ lost per orbit by the
stars (star symbols), cumulative mass $\Delta M_\bullet$ bound
to the SMBH, mass $M$ of the bound star,  apoapsis $d$ of the bound star, periapsis
$r_{\rm per}$ of the bound star.}
  \label{fig:C1plot}
\end{figure}

We find that the periapsis separation of a bound star remains
remarkably constant from one orbit to the next, even when there is
significant mass loss due to Roche lobe overflow at periapse.  As
an example, consider the run C1 in which the initial stars have the
dimensionless Roche lobe parameter $\zeta=0.67$.
As expected for $\zeta < 1$, the bound star does indeed lose mass each
time it sweeps past the black hole.
The gradual decrease of the mass $M$ of the bound star can be seen in the
top panels of Figure~\ref{fig:C1plot}.  The mass
$\delta M$ lost per orbit, shown by the star symbols in the top
panel, increases with each orbit, until after 96 periapsis passages
the star has been completely disrupted.  We note from the
bottom panel that 
the periapsis separation $r_{\rm per}=a(1-e)$ is nearly unchanged 
during this entire process: in this and other cases, we find the bound star returns to the
nearly same relative separation from the black hole regardless of mass
loss.  The apoapsis separation $d=a(1+e)$ is also somewhat constant,
although it decreases at late times when the mass loss is
greatest and strong tidal effects remove energy from the orbit.

\begin{figure*}
  \begin{center}
    \includegraphics[angle=270,width=184mm]{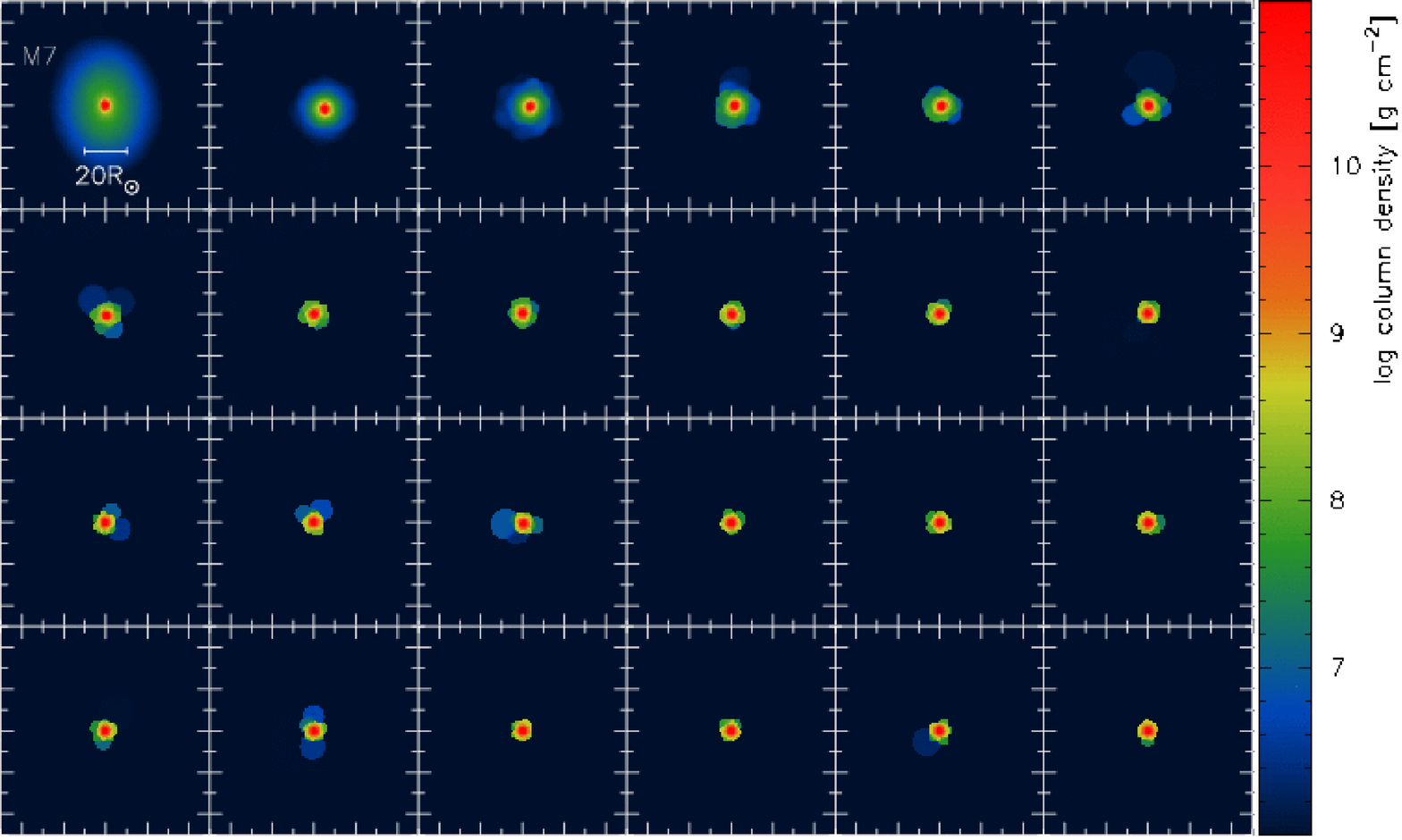}	
  \end{center}
  \caption{Evolution  of the merger remnant formed in run M7 
  after each  periapsis passage when
 the star sets to hydrostatic equilibrium 
 and until mass loss ceases. 
  Time increases from left to right and form top to bottom. 
Initially   the merger remnant has a  large low-density  envelope that is completely removed after
several obits around the SMBH.    }
  \label{fig:multiple-plot}
\end{figure*}

As another example, the lower panels of Figure~\ref{fig:C1plot} 
give the evolution of the merger remnant formed 
in run M5. In this case most of the mass-loss occurs during 
the first periapsis passages where very high entropy material
is removed from the outer layers of the star that responds by reducing its radius. 
In the following evolution, 
mass-loss essentially ceases.
We note that merger products  have a very non-uniform density profile characterized by  a extended low 
density envelope and a dense central region.  Subsequent  passages of the star by the SMBH will 
therefore cause the depletion of the outermost stellar region, unveiling its hot central core. 
An example of this phenomenon is shown in Figure \ref{fig:multiple-plot}, where
we plot column density plots for the merger remnant of run M7.    
After about twenty orbits  mass loss stops and the envelope has been completely removed.
Similar mechanisms, involving tidal stripping suffered by  late-type giants 
during  close passages around a intermediate massive black hole, 
have been invoked in the past \citep{miocchi} to explain the extreme horizontal-branch stars observed  in some globular clusters \citep{Rich}. 

\begin{figure}
  \begin{center}
    \includegraphics[width=84mm]{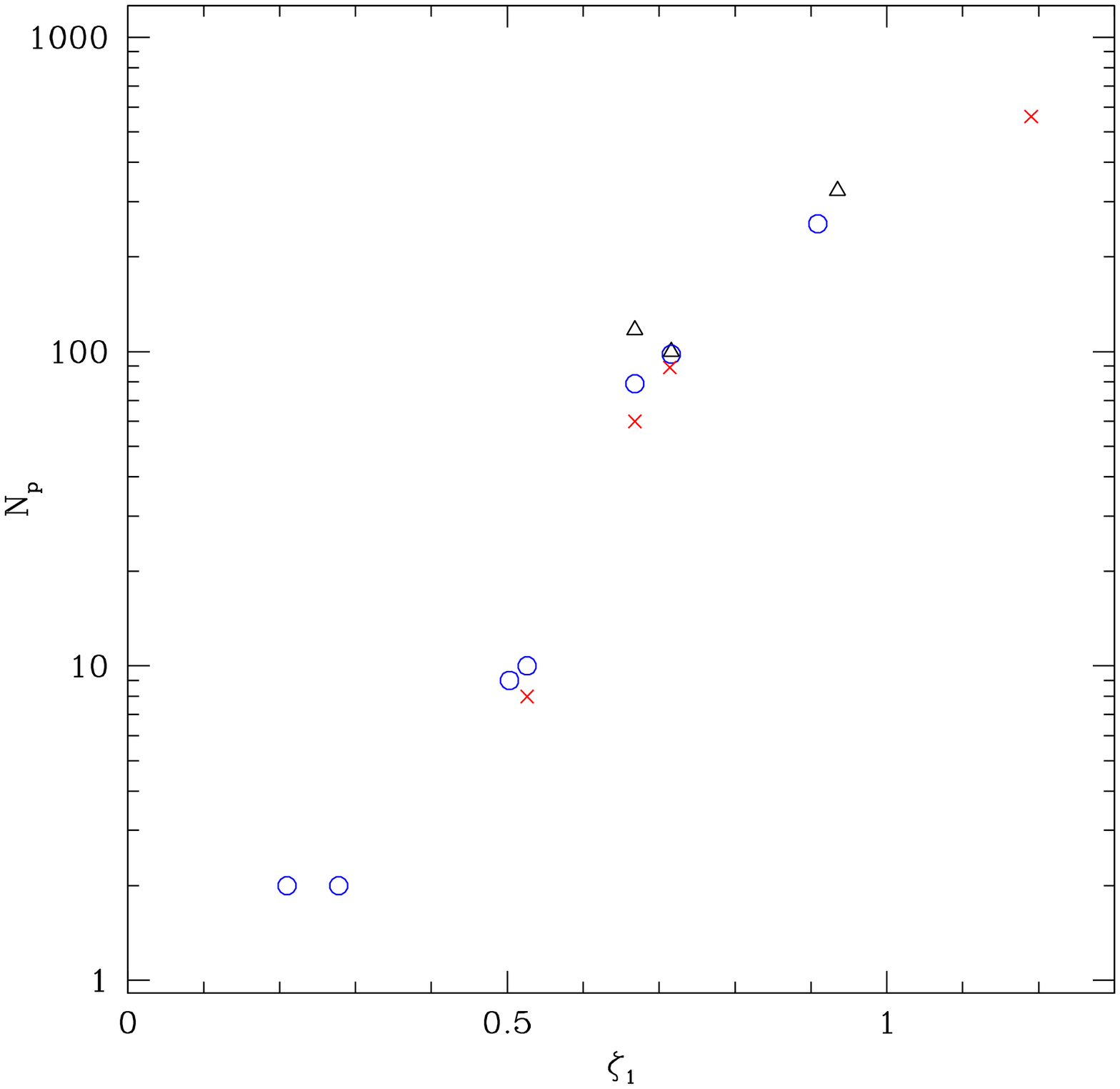}
  \end{center}
  \caption{The number of periapsis passages past the black hole
  needed to
  disrupt the bound star versus the initial dimensionless Roche lobe
  parameter $\zeta_1$.  The different data points represent the
  scenarios in which the bound star suffers a collision (red
  crosses: runs C1, C2, C3, and C4), is formed in a merger (black
  triangles: runs M2, M6, and M10), or is cleanly separated from
  the HVS (blue circles: runs H1, H3, H4, H5, H6, H7, and H8).
  Those collision and merger data with $N_p\gtrsim 100$ likely underestimate $N_p$ due to
  resolution effects and are best considered as lower limits
  (see text).
}
  \label{fig:zeta}
\end{figure}

We find that the dimensionless 
parameter $\zeta$ is strongly correlated with whether and
how quickly a bound star loses mass through Roche
lobe overflow.  For example in run H2, the bound star has $\zeta>1$ and
does not experience a collision or merger that would change $\zeta$: it
consequently continues to orbit the SMBH
without ever suffering any mass loss.
In all the cases with $\zeta<1$, the bound
star is ultimately destroyed after repeated episodes of Roche lobe
overflow, with smaller values of $\zeta$
generally corresponding to fewer orbits before disruption.
Figure \ref{fig:zeta} shows that the number $N_p$ of periapsis passages before
disruption grows exponentially with the initial $\zeta$.  In addition, this
number of passages depends only weakly on whether
the interaction type is a collision (red crosses), merger (black
triangles), or clean ejection of a HVS (blue circles).

The rightmost data point in Figure \ref{fig:zeta}, corresponding to run
C2, deserves some discussion.  In this case, $\zeta_1=\zeta_2=1.19>1$, so that
neither binary component would lose mass if it
were not for the collision induced on the first periapsis passage.
This collision both increases the radius and slightly decreases the mass of the
bound star, effectively decreasing its $\zeta$ parameter to a value below 1.  Thus, on the subsequent passage
past the black hole, the star loses more mass, now due to Roche lobe
overflow.  The response of this particular star to mass loss is that
its radius remains roughly constant.  From equation (\ref{zeta}), the
$\zeta$ parameter then stays below 1 and slowly decreases as the mass ratio $q$
decreases with each successive passage. Ultimately, after nearly 600
periapsis passages, the star is completely pulled apart.

Similarly to the stars from run C2, the $6 \msun$ star in run M13 has
$\zeta=1.22>1$.  This star indeed makes the first periapsis passage
without immediately losing any mass; however, while the binary recedes
away from the SMBH, the merger causes mass ejection.  The resulting
$6.84 \msun$ merger product is large enough that $\zeta$ drops below
1, and, on the subsequent periapsis passages, mass is lost through Roche
lobe overflow.  As a result of shedding its high entropy outer layers,
the merger product shrinks sufficiently that $\zeta$ is pushed back
toward $\zeta\approx 1$.  By comparing runs C2 and M13, we conclude
that the fate of binary stars with $\zeta\gtrsim 1$ depends not simply
on the initial values of $\zeta$ but also on the type of their
interaction and the response of the bound star to mass loss.

In several of our simulations, merger products formed from stars with
$\zeta>1$ are large enough that $\zeta$ drops below 1
and at least some mass is lost
due to Roche lobe overflow on the second and later periapsis passages (runs M1,
M3, M4, M5, M7, M8, M9, M11, M12, and M13).  Because shock heating is
preferentially distributed to the outer layers of a merger product
\citep{2002ApJ...568..939L}, this Roche lobe overflow always strips
away very high entropy material and the product responds by decreasing
its radius.  In this way, the $\zeta$ parameter gradually approaches a
value $\approx 1$, corresponding to an eccentric semidetached binary
consisting of the bound star and the SMBH.  In our SPH simulations of
such cases, we typically follow the dynamics for several hundred
orbits, without seeing an appreciable decrease in the mass of the bound
star: indeed at late times the mass loss typically fluctuates between 0 and 2
SPH particles per periapsis passage.  Such situations necessarily
challenge the mass resolution limit of our simulations, and it is
difficult to say whether such small levels of mass loss are physically meaningful
or simply a numerical artifact.  In any case, in nature, thermal relaxation in
the outermost layers of such a merger product would tend to retract it inside of
its Roche lobe and stabilize the star against further mass loss.

To better understand the effects of the numerical resolution, we vary the 
number of particles used to model several of the scenarios.
The results for scenario M7 are summarized in Table~\ref{rstable}, where
we list the mass, eccentricity, and semimajor axis of the bound star
after 25 orbits.  We find a good agreement of results at all
resolutions tested and a convergence of these results as the number
of particles $N$ is increased up to the value used in this paper
($\approx 4\times10^4$).  In particular, the final orbital data have
converged to within $\sim 0.02$\%.

\begin{figure}
  \begin{center}
    \includegraphics[width=84mm]{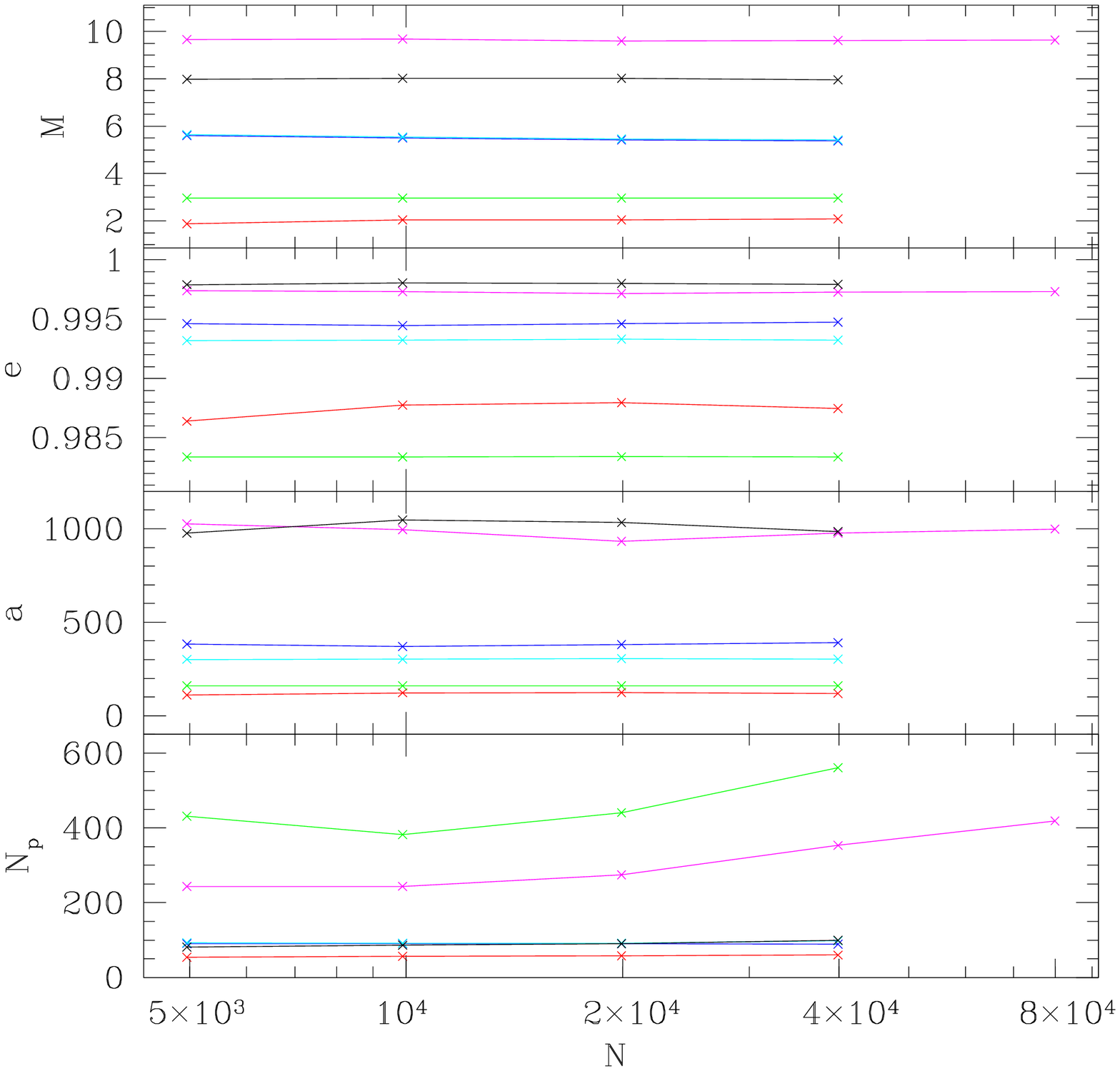}
  \end{center}
  \caption{The mass $M$ in solar masses, eccentricity $e$, and
    semimajor axis $a$ of the bound star, all after 50 periapsis passages, versus total
    particle number $N$ for several representative scenarios in which the bound star
    is ultimately disrupted: C1 (red), C2 (green),
    C4 (blue), H5 (cyan), M6 (magenta), and M10 (black).  Also shown,
    in the bottom frame, is the number $N_p$ of periapsis passages needed
    to completely disrupt the bound star.  For a given scenario, note
    the consistency of the data for $M$, $a$, and $e$ after 50 orbits, as well as for $N_p$ in cases with $N_p\lesssim 100$.
}
  \label{fig:rs}
\end{figure}

We also extend our resolution study to cases in which the bound star is ultimately
disrupted.  We find that, for various particle numbers from $N\approx
5\times 10^3$ up to $8\times 10^4$, the simulations of the same
initial conditions all behave very
similarly for at least the first $\sim 100$ orbits around the SMBH.  
The top three frames of Figure~\ref{fig:rs}, for example, demonstrate
this consistency for the mass $M$,
eccentricity $e$, and semimajor axis $a$ of the bound star after 50 orbits.  For situations in which the
star orbits the SMBH more than $\sim 100$ times, the simulations at
various resolutions diverge at late times, with higher resolutions
simulations in which there is a collision or merger requiring more periapsis passages to disrupt the bound
star (see C2 and M6 the bottom frame of Fig.~\ref{fig:rs}).  
  
In addition to the scenarios shown in Figure~\ref{fig:rs}, we also
study resolution effects in simulations of several other cases in
which the bound star is ultimately disrupted (specifically C3, H1, H3, H4, H6, and M2), with the particle number $N$ varying from $5\times 10^3$ up to $4\times 10^4$.

%
%Interestingly, and in contrast to collision and merger cases with
%large $N_p$, case H3 has the $N_p$ necessary for disruption decrease:
%from 279 for $N\approx 5\times 10^3$ to 255 for $N\approx 4\times
%10^4$.  While the uncertainty in the actual number of passages
%necessary to disrupt the bound star in H3 is large, the trend in data
%does imply disruption in a finite number of orbits.
%
As with the Figure~\ref{fig:rs} data, the consistency of the results
is again very good for $N_p\lesssim 100$.  For example, for case C4,
each of four such simulations predict that it would take somewhere in
the range of 89 to 91 periastron passages to completely disrupt the
bound star.  Furthermore, for case C3 all simulations predict that it
takes 8 periastron passages to disrupt the bound star, in case H4 all
simulations predict that it takes 10 passages, and in case H6 all
simulations predict that it takes 2 passages.  We conclude that the
particle number employed in this paper is sufficient to model
accurately the evolution for at least $\sim 100$ orbits around the SMBH.

\begin{figure*}
  \begin{center}
    \includegraphics[angle=270,width=170mm]{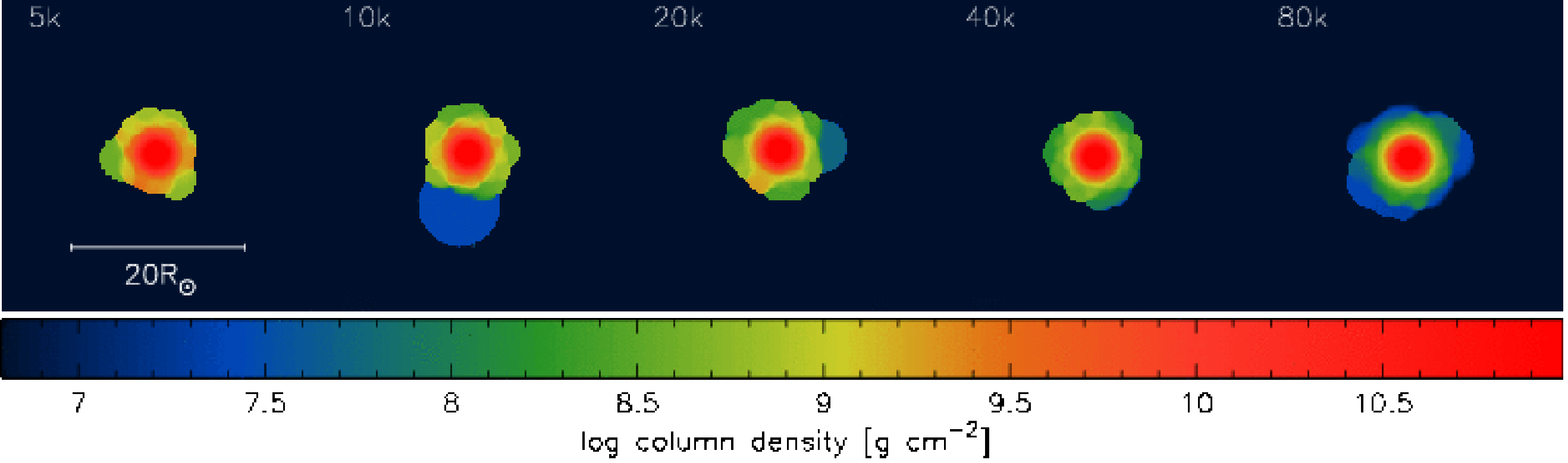}
  \end{center}
  \caption{Column density plots   for run M6 after  50 periapsis passages and using different total number of particles (5k, 10k, ..., 80k).
}  \label{fig:rs2}
\end{figure*}

Finally, as an illustrative example, Figure~\ref{fig:rs2} gives column density plots for run M6 after   50 periapsis passages
and for simulations with different number of particles.

\begin{figure}{~~}
\begin{center}
\includegraphics[angle=0,width=3.2in]{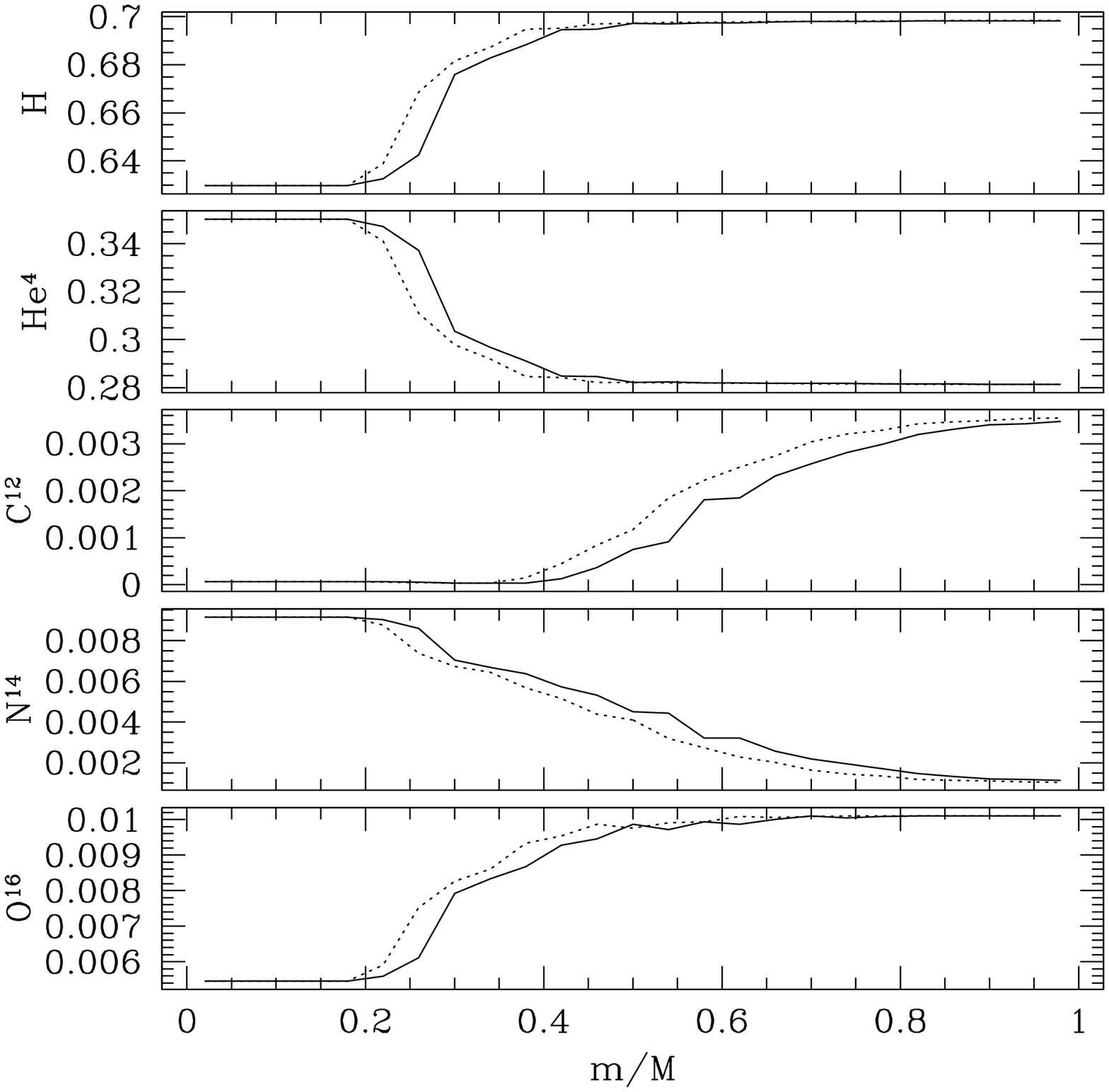}  \\
\includegraphics[angle=0,width=3.2in]{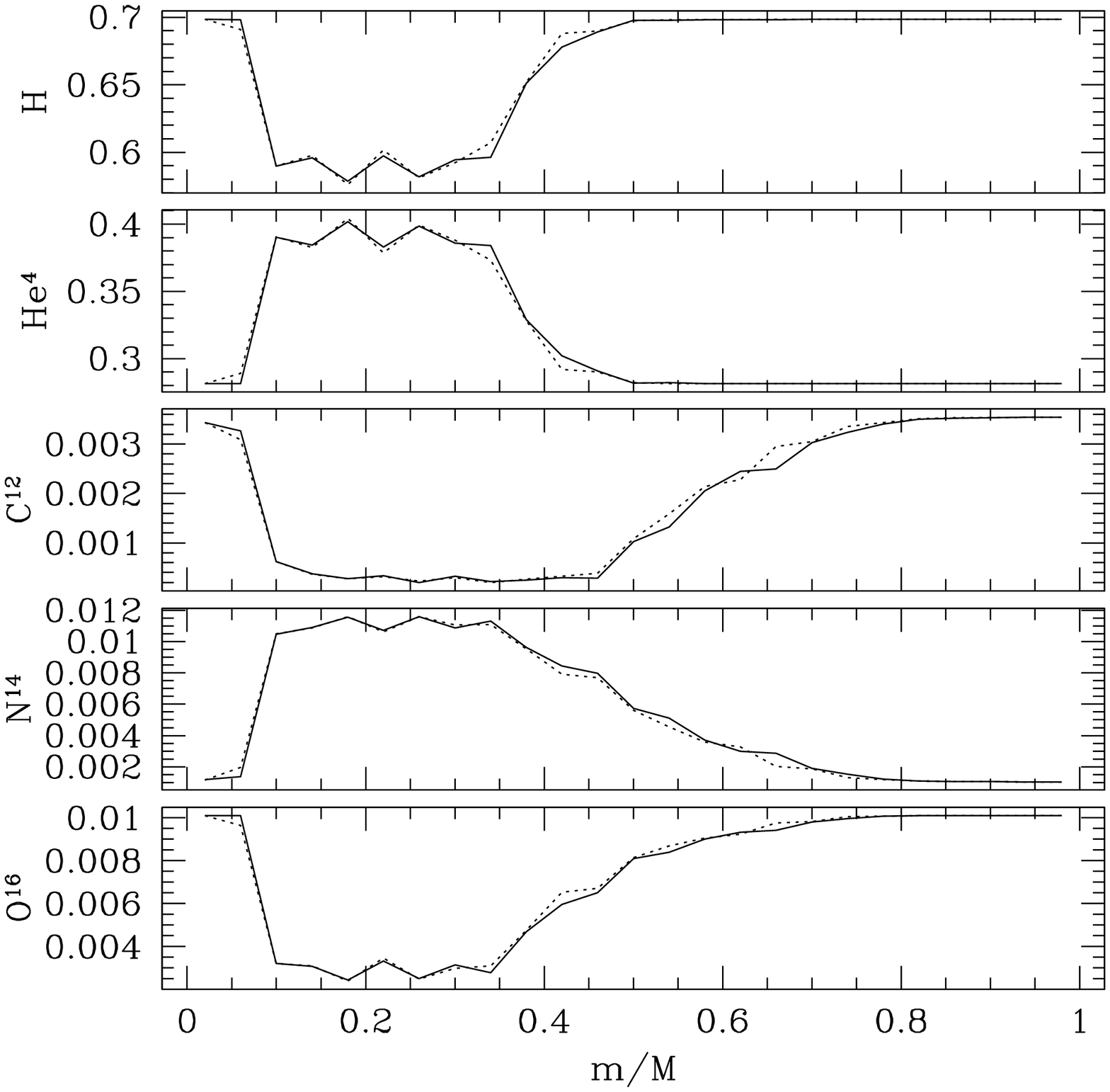} 
\caption{Upper panels: composition profile of the merger product
  in run M4 after two (dotted curves) and eighteen (solid curves)
  periapsis passages, corresponding to masses $M=4.75\msun$ and $M=4.27\msun$, respectively.
  Lower panels: composition profile of the merger product in run M13
  after two (dotted curve) and 76 (solid curves) periapsis passages,
  corresponding to masses $M=6.46 \msun$ and $M=6.32 \msun$, respectively.
}
\label{fig:bf5two}
\end{center}
\end{figure}

\subsubsection{Internal Structure}
Figure \ref{fig:bf5two} shows the composition profiles as a function
of enclosed mass fraction $m/M$ for the bound star in cases M4 and
M13.  Here, $m$ is the mass enclosed within an isodensity surface and
$M$ is the total bound mass.
The dotted curves show the profiles after two periapse passages,
while the solid curves show the same profiles once the bound star has
effectively reached a steady state.  Mass loss experienced during multiple
passages removes the outer layers of the star, decreasing the bound
mass $M$ and causing the
composition profiles to shift slightly to larger enclosed mass
fractions $m/M$.

We note that the helium profile for M13 is qualitatively similar to that of the
case G merger product in \citet{1997ApJ...487..290S}  (see Fig.\ 2 in that
paper): both have a maximum helium abundance at an intermediate radius
inside the star.  In both case G and our M13, the strange helium profile is
caused by a low mass star sinking to the center of the collision
product and displacing the helium rich fluid outward.  The stellar track
for the case G product is shown in Figures 4 and 6 of  \citet{1997ApJ...487..290S}.  In Figure 6, we
see that, on the main sequence, the case G product is somewhat bluer
and brighter than a normal main sequence star of the same mass.  In
Figure 4, we see that the case G product is a little bluer and
brighter than a different collision product (case J) with basically
the same mass but without the dense hydrogen core.  It is the
increased helium content in the stellar interior that
makes the opacity lower (compared to other main sequence stars of the
same mass) and thus bluer and brighter.  So, by analogy, 
our M13 merger product would be a little bluer and brighter
than a normal main-sequence star of the same mass.

The elements Li, Be, and B are potentially interesting observational
indicators of the history of a star.  These elements burn at temperatures of about $2.5\times
10^6$, $3.5\times 10^6$, and $5\times 10^6$ K, respectively, and
therefore can exist only in thin outer layers of the parent stars.  During a
dynamical interaction, these elements can be removed either by
ejecting the outer layers of the star or by redistributing them to an
environment too hot for their long term existence.  Although Be, B,
and Li can exist after the first periapsis passage (see
Fig.~\ref{merg-comp}), the effect of multiple passages is typically to
remove these elements completely.  Although there are cases where B
still exists in the final bound star, it is always severely depleted.
For example, in run M13 the B level at the surface of the final
product is only $\sim$3\% of the surface value in the $6 \msun$
parent star from which it originated.

In Table \ref{t5}, we summarize some properties of the bound stars that survive 
in our SPH simulations (i.e., that are not ultimately disrupted by the SMBH).
 The ``number of passages" represents the number of periapse passages  before 
 the mass loss effectively shuts off, which we defined as having 2 or fewer SPH particles ejected.
The central hydrogen abundance is given by $X_c$, which
always equals the central hydrogen abundance of the lowest
mass parent star.
 We also list  the effective age of the bound star
based on its mass and central hydrogen abundance. 
We evaluate this effective age, based on how long it would take a normal star of that
mass to evolve to that central hydrogen abundance using the TWIN
stellar evolution code.
It is known that the contraction of a merger product to the main-sequence  is very similar to
the contraction of a pre-main-sequence star to the main-sequence.  In the latter case, the
most important variable is the mass (for a given hydrogen
abundance).  In the former case, the two important variables are essentially mass
and  central hydrogen abundance \citep{SL:97}. 
In runs C5 and H2 the bound star is only a slightly perturbed version of one of the
binary components, and the ages in these cases is the same 50 Myr as
that component.  For mergers of two $3\msun$ stars, the effective
age of the merger product is in the range of 14 to 22 Myr.  For
mergers of two $6\msun$ stars, the effective age is in the range
of 6 to 9 Myr.  Mergers of unequal mass stars (M12 and M13) also
significantly rejuvenate a star: for example in M13, the sinking of
the 1 solar mass star to the center of the merger product essentially
resets the nuclear clock to only 0.3Myr after the ZAMS.

In the last three columns of Table \ref{t5} we list  central temperature $T_c$, 
internal energy $U$ and thermal timescale $t_{\rm thermal}$ of  surviving bound stars.
The central temperature  is defined as the temperature in the star where the density is highest and therefore 
is not always the temperature of the highest temperature   SPH particle. 
The central temperature $T_c$
of all the stars is large enough to sustain nuclear burning in the core.
The global thermal time scale can be estimated as
\begin{equation}
t_{\rm thermal}=U/\langle L \rangle,
\end{equation}
where  $\langle L \rangle$ is a mass weighted average of the
luminosity $L$ throughout the entire star.

To calculate $L$, we take advantage of the fact that the parent stars are massive enough to be fully radiative.  
In addition, shock
heating prevents any convective zones from
existing in a newly formed merger product.
  Thus, we obtain the luminosity $L$ exiting a closed surface
by the integral $ L=\oint{\bf F}\cdot {\bf da}$, where ${\bf da}$ is
an area element on the surface and 
the diffusive radiative flux
${\bf F}=-4acT^3{\bf \nabla}T/(3 \kappa\rho).$
Here $a$ is the radiation constant, $c$ is the speed of light, and $\kappa$ is the opacity.
The surface integral is easily converted to a volume integral by the
divergence theorem.  The result, $L=\int {\bf \nabla}\cdot {\bf F} dV$,
is straightforward to estimate in SPH:
\begin{equation}
L=\sum_i \frac{m_i}{\rho_i}({\bf \nabla}\cdot {\bf F})_i, \label{luminosity}
\end{equation}
where the sum is over only those particles positioned inside the
surface under consideration.  Because SPH calculations
cannot properly resolve the photosphere, equation (\ref{luminosity}) cannot
be used to give a reliable total luminosity.  However,
equation (\ref{luminosity}) does allow us to study the luminosity profile
throughout the bulk of the system.

\begin{figure}{~~}
\begin{center}
\includegraphics[angle=0,width=3.2in]{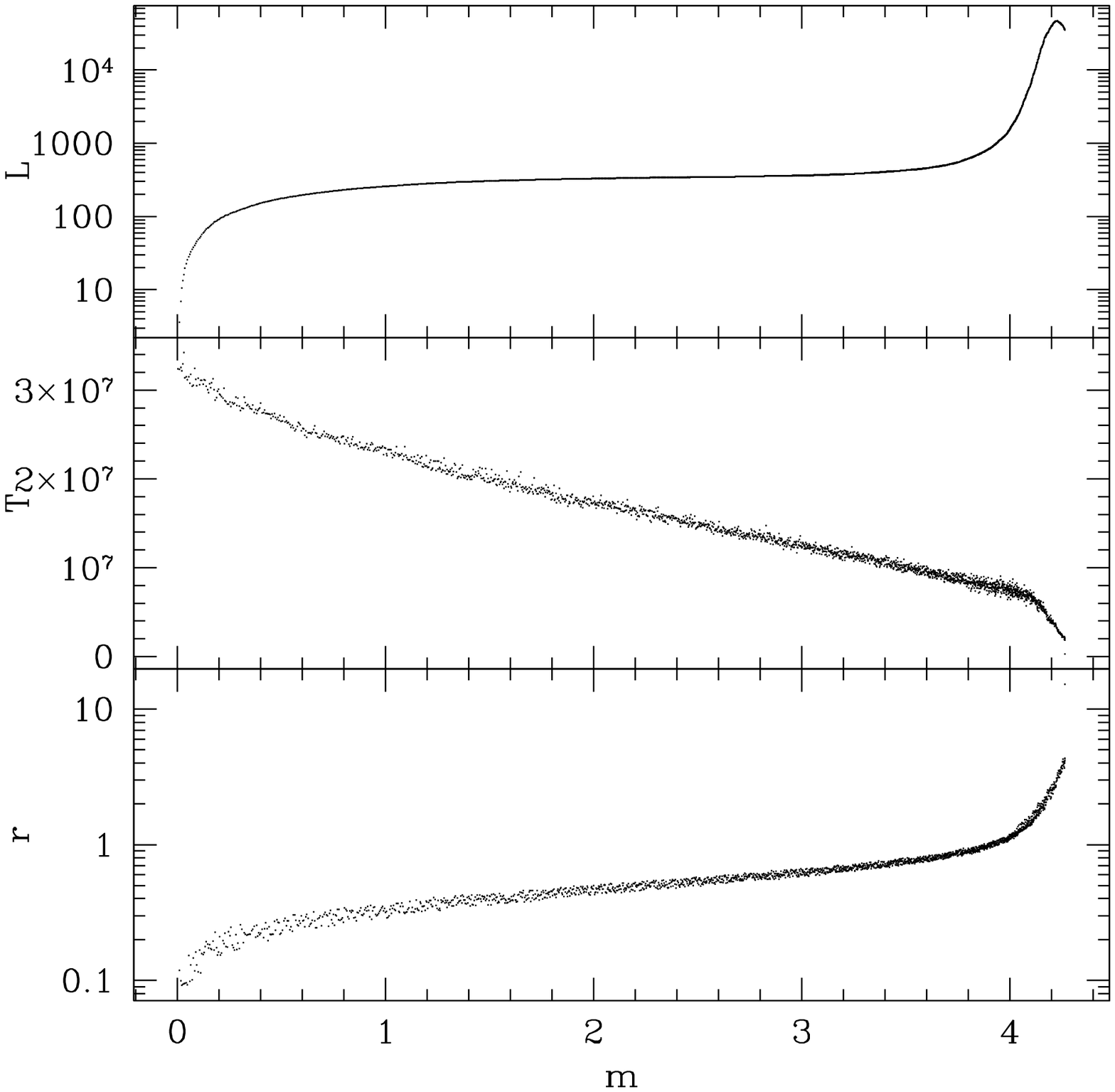}  \\
\includegraphics[angle=0,width=3.2in]{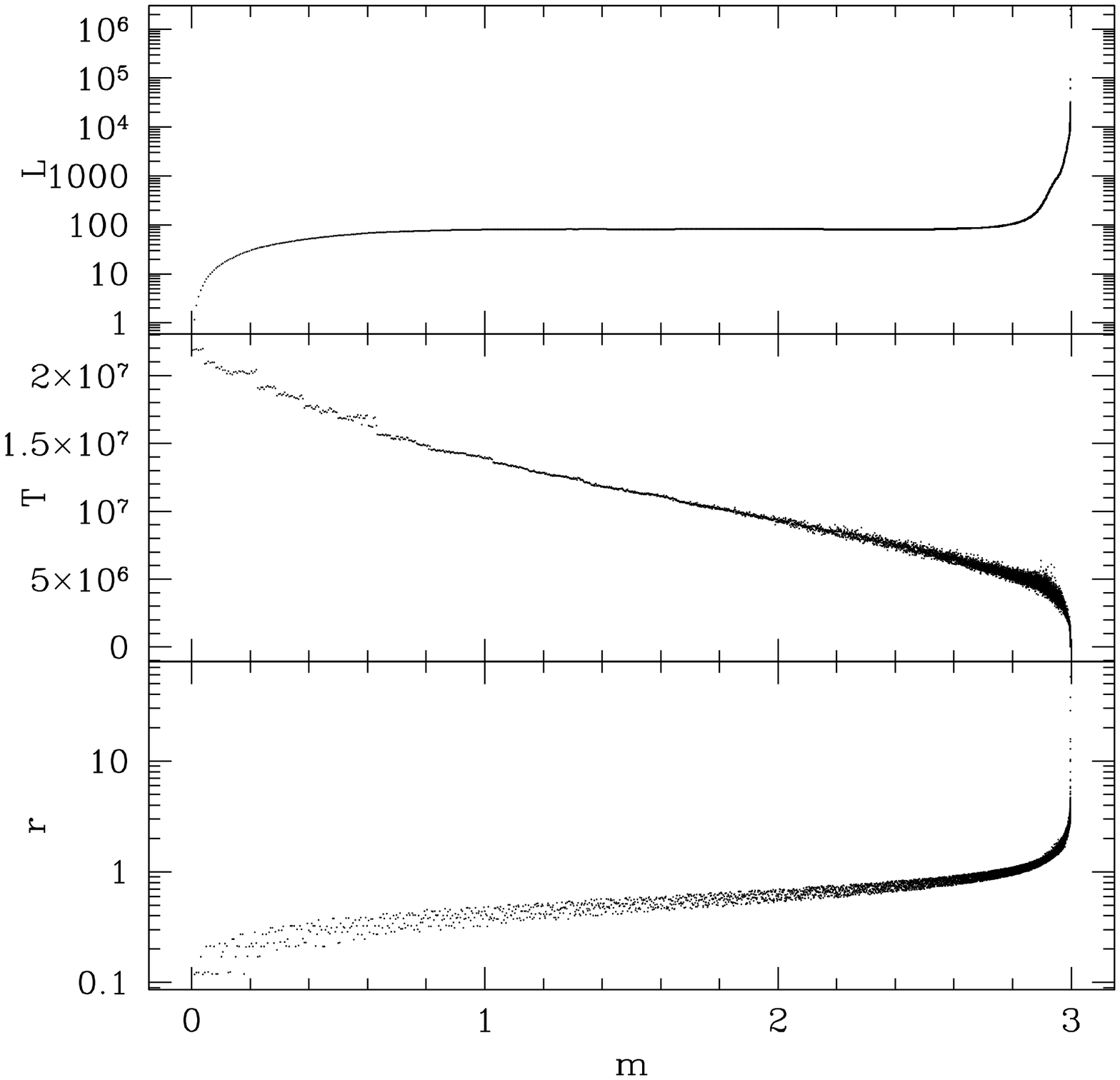} 
\caption{From  top to   bottom: luminosity $L$, temperature $T$, and radius $r$
as a function of  enclosed mass $m$ for the final bound stars of runs M4 (upper panels)
and C5 (lower panels). $L$, $r$ and $m$ are  in solar units, while $T$ is in Kelvin.
}
\label{int}
\end{center}
\end{figure}

As an example, in Figure~\ref{int}, we show a  more detailed look at
the interior of the final bound stars
in runs M4 and C5. From top to bottom, we give the luminosity $L$, temperature $T$, and radius $r$
as a function of  enclosed mass $m$.   To evaluate the luminosity
profile, we use equation (\ref{luminosity}) on each SPH particle, summing
over particles of larger density and calculating the opacity $\kappa$ from the OPAL tables.
Our merger and collision products typically achieve a maximum luminosity in their
outermost layers that is comparable to the Eddington
luminosity
\begin{equation}
L_{\rm edd}=\frac{4\pi G c}{\kappa} M \sim 3.8\times10^4 {\rm L_{\rm \odot}} \frac{M}{\msun}~~,
\end{equation}
although such a high luminosity would diminish rapidly as the star
contracts to the main-sequence in a time $t_{\rm thermal}$ usually of order $\sim 0.1 {\rm Myr}$.

\section{Discussion}
The observed rotation rates of HVSs may give important clues to their formational history.
\citet{HA:07} has proposed that,
as a consequence of tidal locking  in close binaries, 
HVSs ejected by the Hills mechanism should rotate systematically slower than 
field stars of similar spectral type.
\citet{MB:08} found that the late B-type star HVS 8 has a rotational velocity of
$\sim 260 \kms$, more typical of single B-type stars and therefore seemingly
contrary to the hypothesis 
of a binary  origin for this star.
 In order to explain the observations, other ejection mechanisms have been invoked, such as
 ejection by a close encounter with a massive black hole binary or with a stellar black hole
 orbiting the galactic center SMBH \citep{YU:03, L:06, S:06, LB:08}.  
However, it has been note  that a larger statistic would be certainly required in order for the rotation to be used
as a signature for the origin of HVSs and/or S-stars \citep{PER:09}. 
Furthermore, in this paper we have shown that there are two other  potentially important ways with which the stars 
 can somewhat increase their  rotation  even in the binary disruption scenario: tidal torque by the  SMBH  at periapsis (if the stars enter within 
 their tidal disruption radius)  and/or a collision between the two binary members.

\begin{table*}\scriptsize 
\caption{ \label{t4}} 
{Same as table \ref{t2} but for simulations in which the stars do not collide and one member becomes a HVS.
$J_{\rm {captured}}$ gives the spin of the stars that remain bound to the SMBH, while $J_{\rm {ejected}}$ refers to the ejected stars.
}
\begin{center} 
\begin{tabular}{lllllllr}
 \hline
 ${\rm Run}$  & $v_{\rm ej}$ &      a                      &  e   & $\Delta M_{\rm b}/M_{\rm b}$              &  $\Delta M_{\bullet}/M_{\rm b}$ &          $J_{\rm {captured}}(J_{\rm{ejected}})$    \\
                         &     (km/s)       &     (au)                  &           &       &             	         			    &                                                 	         &	                            \\
 \hline
 H1      &  3883(3910)&  160(160)    & 0.991(0.991) & $-2.230\times 10^{-4}$         &  $1.221\times 10^{-4}$    & $0.211 \times 10^{-2}  (0.188 \times 10^{-2}) $ \\
 H2      &  2026(2160)&  309(308)    & 0.983(0.974) & $0$          &  $0$    &   $  0.268\times10^{-3} (0.238\times10^{-3})$  \\
 H3      &  1082(1100)&  423(421)    & 0.991(0.983) & $0$         &   $0$    &    $      0.153\times10^{-3} (0.729 \times10^{-4})$                 	\\
 H4      &  4221(4252)&  142(142)    & 0.995(0.990) & $-1.019\times 10^{-2}$         &  $5.401\times 10^{-3}$ &  $ 0.194 \times 10^{-1} (0.147\times 10^{-1})$\\
 H5      &  2121(2155)&  306(307)    & 0.974(0.993) & $-8.454\times 10^{-5}$         &  $4.759\times 10^{-5}$   &  $0.149\times 10^{-2}  (0.168\times 10^{-2}$)\\
 H6      &  1637(947.6)& 174(444)    & 0.988(0.999)         & $-0.9914$          & $0.9914$  &  $0.445\times 10^{-1}  (0.379 \times 10^{-1})$    \\
 H7      &  2570(2574)&  677(674)    & 0.983(0.978)  & $-1.335\times 10^{-2}$          &  $6.911\times 10^{-3}$     &   $0.223\times 10^{-1}  (0.348\times 10^{-3}$)  \\
 H8      &  2494(2497)&  619(619)    & 0.991(0.990) & $-0.4835$                                 &  $0.2458$    &       $0.845 \times 10^{-1} ( 0.274\times 10^{-1})$   \\
\hline
\end{tabular}
\end{center}
\end{table*}

With the help of simplifying approximations, we are able to relate the
rotational parameter $J$ calculated in our simulations to the
observable rotational velocity $v$ after that star has thermally
relaxed back to the main sequence.  In particular, we approximate that
the star rotates rigidly, that its rotation does not drastically
affect its structure, and that the rotational parameter $J$ is
conserved during relaxation.  Using $L=Iv/R=c_1 M R v$ and
$E=-c_2GM^2/R$ in equation (\ref{PP}), we obtain
$J=c_1c_2^{1/2}v(R/(GM))^{1/2}.$ Clearly $c_1$ and $c_2$ are simply
numerical coefficients related to the moment of inertia $I$ and total
energy $E$ of a star, respectively.  For B-type main sequence stars, we
find $c_1c_2^{1/2}\sim 0.04$ to
$0.05$ and $R/M\sim$0.5 to 0.8 ${\rm R_\odot/M_\odot}$ using models from the
TWIN stellar evolution code.  Solving for $v$ in terms of $J$, we find
\begin{equation}
v\sim 1.2\times 10^4 J\,{\rm km\,s}^{-1},\label{vfromJ}
\end{equation}
accurate to within $\sim30$\% for most B-type main sequence
stars.  Given the $J$ values of ejected stars in our simulations (see
Fig.~\ref{rotC} and Table \ref{t4}), we estimate from equation (\ref{vfromJ}) that the post-relaxation
rotational velocity $v$ can be as large as $\sim$400 or 500 km s$^{-1}$ for
HVS stars (consider runs C3 and H6).  We therefore conclude that the rotation of the star HVS 8, for example,
is completely consistent with a binary origin.

\begin{table}\scriptsize
\caption{\label{rstable}}{Resolution study for scenario M7.  The particle number is
  given by $N$, while the mass, orbital eccentricity and
  semimajor-axis of the bound star after 25 orbits around the SMBH are given by $M$, $e$, and $a$ respectively.
}
\begin{center}
\begin{tabular}{rlll}
\hline
$N$ & $M$      & $e$ & $a$\\

    & ($\msun$)&     &(au)\\
\hline
4,957 &0.9972 &1147.1 &8.468\\
9,889 &0.9973 &1146.7 &8.410\\
19,933 &0.9974 &1146.2 &8.392\\
39,877 &0.9973 &1146.2 &8.390\\
\hline
\end{tabular}
\end{center}
\end{table}

The $J$ values for unmerged stars in our simulations that are bound to the
SMBH after the first periapsis passage indicate, via equation
(\ref{vfromJ}), that the post-relaxation rotational velocity $v$ would be typically
$\lesssim 400$km s$^{-1}$ but could be as large as $\sim 1000$km
s$^{-1}$.  These large rotation velocities, however, correspond to
stars that penetrate deeply within their tidal disruption radius
(e.g.\ runs H6 and H8) and therefore are eventually destroyed after
several orbits.
Merger products obtain even larger spins after the first periapsis passage
(see Table \ref{t3}).  
However, as the merger product keeps orbiting the SMBH,
$J$ decreases as mass gets pulled off 
the outside of the star, where the specific angular momentum is greatest.
The bound stars that survive to orbit the SMBH end
our simulations with $J\lesssim 0.017$ (see Table \ref{t5}),
corresponding to post-relaxation rotational velocities $v\lesssim
200$km s$^{-1}$.  As our simulations began with irrotational stars,
the actual final rotational velocity of a (bound) star could be 
larger or smaller if the parent stars had significant spin, depending on the orientation of the spin axis
with respect to the external orbital plane and/or the plane on which the  collision occurs.
If the spin axis is aligned with the angular momentum of the external orbit, the initial spin will sum up 
with that acquired due to the tidal torque from the SMBH. A larger spin will also result  
from a collision, if the angular momentum of the inner binary is initially aligned with the spin axis of the stars. 
We stress here that, in general, the effect of an initial spin on the final rotation of the stars can be complicated, for this reason we 
decided to ignore initial rotation and begin with binary components irrotational in the inertial 
frame, which allows us to more easily measure any rotation imparted during the subsequent interaction.

Deep near-IR  observations  of the GC show that the S-stars are B0-B9 main-sequence stars
with rotational velocities similar to those of field stars of the same spectral type \citep{A:05}.
Our final bound stars therefore have properties very similar to those of the S-stars:
their masses qualify them as spectral type B main-sequence stars,
and their post-relaxation rotational speeds are of the correct general magnitude.
For example, the rotational speeds of our fastest rotators are consistent with the 
$220 \pm 40$ km s$^{-1}$ value for the S-star SO-2 \citep{2003ApJ...586L.127G}.
However, we note that the orbital eccentricities of our bound stars ($0.96<e<1$) are
larger than that of SO-2 ($e \approx 0.87$), as similarly found in simulations by \citet{GL:06} and \citet{HA:07}.

\begin{table*}\scriptsize
\caption{ \label{t5}} 
{Some properties  of the bound stars that survive 
in our SPH simulations.  ``Number   of Passages'' gives the number of SMBH-star encounters
before mass-loss ceases. Mass, orbital eccentricity and semimajor-axis of the star are given by $M$, $e$ and $a$ respectively.
The value of $J$ is the final dimensionless spin parameter, while $X_c$  is the central hydrogen abundance. 
The corresponding effective age is also listed.
In the last three columns,we give central temperature $T_c$ (the temperature where the density is highest),   
internal energy $U$, and thermal timescale $t_{\rm thermal}$.
}
\begin{center}
\begin{tabular}{lclllllllll}
 \hline
Run & Number   of Passages &  $M$     & $e$       &  $a$ & $J$   &  $X_c$ & Effective Age & $T_c$ & $U$  &   $t_{\rm thermal}$ \\
        &                                          &    ($\msun$)&             &  (au)&          &               &      (Myr) & $\times10^6$ (K) & $\times10^{50}$ (erg) & (Myr) \\
        \hline
C5       &    8   &2.998  &    0.974      & 212 & 0.009 &0.63 &50 &  22         & 0.141	& 0.4	        		\\
H2        &   2   &3.000     & 0.978      & 316  & 0.000 &0.63 &50 & 23         & 0.148     & 1     		\\
M1        &  30 &  4.83      & 0.995     &  1010& 0.009 &0.63 &16 & 31	& 0.326 	&  0.1		\\
M3        &  30  & 4.94      & 0.996      & 996  &0.010 &0.63 &16	& 31 	& 0.321	& 0.2			\\
M4        &  18   &4.27      & 0.996       &1000 &0.009 &0.63 &22	& 32	& 0.292	& 0.2			\\
M5        &  22   &5.15      & 0.962      & 1020 &0.014 &0.63 &14	& 32	& 0.309	& 0.05		\\
M7        &  25   &8.39      & 0.997      & 973  &0.016 &0.56  &9	& 39	& 0.710	& 0.1			\\
M8        &  90   &10.7       &0.996     &  1030 &0.014 &0.56  &6	& 39	& 0.944	& 0.07		\\
M9        &  27   &8.74       &0.995     &  1010 &0.010 & 0.56  &9	& 39   	& 0.736	& 0.06		\\
M11      &   14   &10.1      & 0.980      & 1010 &0.005 &0.56  &7	& 36	& 0.793	& 0.03		\\
M12      &   27   &7.27      & 0.992    &   1020 &0.009 &0.63  &7	& 32	& 0.269	& 0.03		\\
M13      &   79   &6.32      & 0.997 &      1060 &0.017 &0.70  &0.3 & 18	& 0.391	& 0.1			\\
 \hline
\end{tabular}
\end{center}
\end{table*}

%Also, an S-star generated by a stellar merger/collision would be expected to rotate faster than a star
%formed in a tight binary.

For tidal torque from the black hole to have a significant effect on stellar rotation,
the stars should enter deep into their  disruption zone (i.e., $r_{\rm per}<r_{\rm t}$).

When no collision occurs between the components of a binary,
the distance of closest approach of the two stars to the SMBH
typically changes little due to the encounter. 
In such cases, a necessary condition for significant spin-up
is that the binary itself be on an orbit that passes within
$\sim r_{\rm t}$ of the SMBH.
This implies in turn that the fractional change in orbital angular
momentum with respect to the SMBH, per orbit, be of order unity.
This condition is satisfied in the so-called ``full loss cone''
regime, which, in a galaxy like the Milky Way, extends inward
to $\sim 0.2$ times the SMBH influence radius,
or to $r\approx 0.5 {\rm pc}$  (e.g.\ Wang \& Merritt 2004).
\footnote{Assuming a $\rho\sim r^{-2}$ density cusp.}
Inside this region, which is the region of interest for the
current study, evolution onto loss-cone orbits is diffusive,
and most binaries would be tidally disrupted before finding
themselves on orbits that intersect $\sim r_{\rm t}$.
We note however that in the ``massive-perturber scenario" the apoapsis
distance of the binary is of the order of a parsec (i.e., $> 0.5{\rm pc}$) and therefore  the fractional 
change in orbital angular momentum with respect to the SMBH, per orbit,
can be of sufficient  to put the binaries on  a  trajectory  that passes within  $r_{\rm t}$ of the SMBH.
Even inside $0.5$pc, other dynamical processes like resonant relaxation \citep{RT:96}, scattering from an 
intermediate mass black hole \citep{MA:09}  or
perhaps eccentric instability in a disc \citep{MAD:08} can produce larger changes in orbital angular momentum
than in the case of two-body relaxation alone. 
On the other hand tidal spin-up  is not expected to be very efficient because  it is important only for 
the  narrow range of periapses: $\frac{1}{2}r_{t} \lesssim r \lesssim r_{t}$ (for $r \lesssim  \frac{1}{2}r_{t} $ the star is fully 
disrupted; for  $r >r_{t}$ tidal torque  is  small).
As a consequence of the previous condition,  the ejected star will  lose a large fraction of its mass.
An observational indicator of the history of the star would be, even in this case,   a deficit 
in the abundances of  light elements (such as Lithium) that can exist only in thin outer layers 
of the parent star and that are typically removed  by the tidal interaction with the SMBH.

One of the main arguments against rejuvenation of the S-stars through merger 
is the apparent ``normality" of their spectra \citep{fig}.
We note, however, that the envelope of our merger products will not look significantly  different than
that of normal stars (compare the right edge of the plots in Figure 15 
with the right edge of Figure 2).  In fact, if the
parent stars are of equal mass, the merger product will have very
normal profiles throughout the star.  If the parent stars are of
significantly different mass, then the profiles are more peculiar and one
should worry about how this affects the stellar evolution.  
The main effect would probably be  to change the opacity and therefore  shift slightly 
the color and luminosity. But, unless significant mixing is induced,
the chemical composition of the outermost envelope will remain similar to that of 
the higher mass star (with the possible exception of Li, Be, and B levels: see \S\ref{MG}).

The tidal disruption of  a star  passing close enough by a SMBH to enter its tidal radius, 
produces a luminous UV/X-ray flare of radiation 
as the bound stellar gas falls back onto the black hole and is accreted \citep{R:88}.  
Tidal flares are of great interest because they can probe the presence of SMBHs in 
galaxies with otherwise no evidence of an active nucleus and can be used 
to measure  the mass and spin of the  central  black hole \citep{KM:04,GZ:08,GZ:09}.
Computations of the tidal disruption of stars 
have been performed by several studies in the past, with the aim of
understanding the observational signatures of these events \citep{U:99,BG:04,GC:05,LD:09,SQ:09,GL:09,KR:10}.
We note that there are  many important scenarios in our paper that 
have been  so far almost completely  ignored: (i) multiple passages by the same star;
(ii) merger of two stars resulting in disruption due to the increased
size; (iii) partial tidal disruptions.
   
Predicting the  radiative effects of multiple passages of a star by a SMBH is outside 
the scope of the present work.
But, it seems likely   that the light curve resulting from these repeated
tidal events might show a series of small peaks, separated roughly
by the orbital period, before finally producing the large peak that is
observed as the ``tidal disruption."
Assuming the star on a parabolic trajectory, after the star-SMBH encounter, 
the most bound material moves on a orbit with semi-major axis $a=\frac{1}{2}r^2_{\rm per}/R$
and returns to periapsis after a time
\begin{eqnarray}
t_0=\frac{2\pi r_{\rm per}^3}{\left(GM_{\bullet} \right)^{1/2} (2R)^{3/2} }= 0.22  \nonumber  ~~~~~~~~~~~~~~~~~~~~~~~~~\\
\times \left( \frac{r_{\rm per}}{r_{\rm t} } \right)^3 \left( \frac{R}{\rm R_{\odot}} \right)^{3/2} 
\left( \frac{M}{\msun} \right)^{-1}
\left(\frac{M_{\bullet}}{4 \times 10^6 \msun}\right)^{1/2} {\rm yr}~,~~~~
 \end{eqnarray}
 that it is also the time when the  flare starts, while the peak return rate occurs at $t \sim 1.5 t_0$ \citep{EK:89,L:02}.

 As previously discussed,  merger products are very large, so it is easy to strip
off lots of mass during the early periapsis passages, while at later time 
the mass loss often ceases. The light curve resulting from these repeated
tidal events will eventually  show a series of small peaks of declining  intensity (see Figure 12).
The result of the  repeated (partial-)tidal disruption of a star with a large envelope (e.g., late-type giants) will 
show a similar  light curve.
In collisions without mergers, there is some expansion in size but it
is not as
dramatic as in a merger.  So it is not until late times that there is
significant mass overflowing the Roche lobe. The light curve will show peaks of increasing intensity until the
last brightest flare produced by the full tidal disruption  of the star.

In future work, one could model the material that becomes
bound to the black hole more carefully.  In particular, it would be interesting to identify
possible signatures of interactions between the bound star and the
accretion torus left behind from previous periapse passages.
Quasi-periodic  emission  may be detected if X-ray flares arise every time the  star crosses the torus plane \citep{DFB:10}.

\section{Summary}
In this paper, we carried out hydrodynamic simulations of binary stars in orbit
about the supermassive black hole (SMBH) at the Galactic center.
In the $N$-body simulations of Paper I,
we assigned physical sizes to stars based on a simple mass-radius relation
and predicted which binaries would merge, i.e., undergo a collision with relative velocity less than escape velocity.
The fluid simulations presented in this paper were found to be quite consistent with the
$N$-body simulations,
in the sense that when the latter predicted a stellar merger, the fluid stars typically merged as well.
The merger rates presented in that paper are therefore confirmed by the present work.
The principal, new findings of our work are summarized here.
\begin{itemize}
\item[1] The central temperature of the merged stars in all our simulations is large
enough that there would still be nuclear burning in the stellar core.
However, mergers tend to ``rejuvenate'' stars, in the sense that the lower mass star sinks
to the center of the merger remnant, effectively resetting the nuclear clock of the merger
product to the zero-age main sequence. Even the products of equal-mass mergers exhibit
significant rejunvenation.
\item[2] Mass loss during collisions is generally small, but large fractional
mass loss can occur when one or both stars is tidally perturbed by the SMBH;
if the two stars merge, a temporarily more extended object is formed, further
enhancing the mass loss rate.
When the merger product has a distance of closest approach to the SMBH
smaller than its Roche limit, total disruption always occurs after repeated periapse
passages.
Repeated tidal flares, separated by roughly the orbital period, are predicted
to precede the disruption.
\item[3] In stellar mergers, elements that can exist only in the outermost envelope
of the stars such as Li, Be and B are severely depleted, due primarily to tidal 
truncation by the SMBH in subsequent periapse passages.
However, the envelopes of the merger products do not otherwise differ significantly from
those of the parent stars.
\end{itemize}

We finally stress that SPH calculations neglect radiative and heat transport, and therefore can follow the system
only over hydrodynamical timescales that are typically of the order of a few hours. For these reasons, in this paper,
we were able to discuss the relaxed structure of merger products only qualitatively and the relaxation time only in order of magnitude. In a subsequent paper, we plan to present  the results of stellar evolution calculations that can follow the evolution of the SPH merger products over much longer, thermal and nuclear timescales and determine their track in a color-magnitude diagram. These calculations will allow us to compare the observable properties of our models with the properties of stars observed at the galactic center.

\bigskip\bigskip

\acknowledgments
We thank E.~Gaburov for his development of the GPU library used to
calculate gravitational forces and energies in this paper,
and S.~Mikkola who wrote the ARCHAIN code.
We also thank H.~Perets, W.~R.~Brown, J.~Faber, A.~Gualandris,
N.~Ivanova, S.~J.~Kenyon, S.~Komossa, S.~Portegies Zwart,
and A.~Robinson for helpful discussions.
Column density plots in this paper were produced
using the publicly available visualisation tool SPLASH \citep{PC:07}.
This work was supported by grants NNX10AF84G and NNX07AH15G from NASA
and by grants AST-0807910 and AST-0821141 from the NSF.

\appendix

\section{Unequal mass binaries: $N$-body simulations}
In the following, we briefly present the results of new $N$-body simulations of unequal-mass binaries passing by Sgr A*.
The integrations are performed by using  the high accuracy $N$-body code {\tt ARCHAIN} \citep{MA:02,MM:06}.  
The  binaries are  initially placed at a distance $d=0.1-0.01$pc from the SMBH with
a purely tangential velocity corresponding to periapses between the tidal disruption radius of the secondary star 
  and $r_{\rm bt}$.  The primary star has mass $M_1=6{\msun}$, while the mass of the secondary  
 is either $M_2=1{\rm M_{\odot}}$ or $3{\rm M_{\odot}}$.  We adopt
 $a_0=0.1$au for the internal semimajor-axis of the binaries.
 In all the simulations the final integration time is fixed to one orbital period of the binary orbit around the SMBH.
 In total we perform 1200 simulations.
 
As expected,  we find a systematically larger ejection velocity for the less massive star \citep{YU:03}.
For binaries with $M_2=1{\rm M_{\odot}}$($3{\rm M_{\odot}}$)   and initial distance $d=0.01$pc, the mean asymptotic ejection velocity 
of the primary and secondary stars are respectively: 
$v_1\sim 1000{\rm km\,s^{-1}}$ ($1600{\rm km\,s^{-1}}$) and   $v_2 \sim 2800 { \rm km\,s^{-1}}$ ($ 2700{\rm km\,s^{-1}}$).
When   the initial apoapsis of the binary is increased to $d=0.1$pc we found:
 $v_1\sim 1350{\rm km\,s^{-1}}$ ($2180 {\rm km\,s^{-1}}$) and   $v_2 \sim 3500 { \rm km\,s^{-1}}$ ($ 3200{\rm km\,s^{-1}}$).
Table \ref{ta} gives the fraction of collisions,
mergers, and HVSs ($v_{\rm ej}>1000{\rm km\,s^{-1}}$) in the simulations.
Note that in this table  any merger is also counted as a collision.
In the table we report the probability of ejection, distinguishing between  the two components of the binary.
It is clear that the initial distance of the binary from the central black hole plays a fundamental role in 
determining which member is ejected.
For large initial distances (i.e., $d=0.1$pc) the ejection probability is almost independent on the 
stellar mass, while for $d=0.01$pc, the lighter star is preferentially ejected. 
These results are consistent with 
the findings of \citet{Sari:2010} that used an approximated method to study the dynamical evolution
of binaries on parabolic orbits.  
Our simulations suggest that, in the limit that the external orbital energy of the binary goes to zero,  
the ejection probability becomes  an independent function of the stellar mass.

\begin{table*}\scriptsize
\begin{center}
\caption{HVSs ($v_{\rm ej}>1000\kms$), Collision and Merger frequency(\%) \label{ta}} 
\begin{tabular}{lllllll}
 \hline
 $M_2$(${\rm M_{\odot} }$)  & $d$(pc)	& HVSs (total)  & HVSs (primary)  &  HVSs (secondary) & Collisions &  Mergers   \\ 
\hline
 1       	&  0.01	&  33. 6	 &    1.29 	&    32.3   &  7.74  & 7.10   \\
 3               &  0.01     & 52.9          &    25.5   &    27.4   &  9.35  & 8.71  \\
 1	         &  0.1        &  62.5        &   31.3            &       31.2        &6.77      & 6.45 \\
 3	         &  0.1        &   59.3       &29.3       & 30.0         &11.6     & 11.3         \\
\hline
\end{tabular}
\end{center}
\end{table*}

\end{document}